\documentclass[12pt,a4paper,ngerman]{article}

\usepackage{graphicx}
\usepackage{dcolumn}
\usepackage{bm}

\usepackage[T1]{fontenc}
\usepackage{mathptmx}
\usepackage{float}
\usepackage{xcolor}
\usepackage[superscript]{cite}
\usepackage{braket}
\usepackage[a4paper, left=3cm, right=3cm, top=3cm]{geometry}
\usepackage{ragged2e}
\usepackage[onehalfspacing]{setspace}
\usepackage{tabularx}

\begin{document}

\newcommand{\ba}{\begin{eqnarray}}
\newcommand{\ea}{\end{eqnarray}}
\newcommand{\bc}{\begin{center}}
\newcommand{\ec}{\end{center}}
\newcommand{\be}{\begin{equation}}
\newcommand{\ee}{\end{equation}}
\newcommand{\EA}{\emph{et al.}\xspace}
\newcommand{\ra}{\rightarrow}
\newcommand{\bi}{\begin{itemize}}
\newcommand{\ei}{\end{itemize}}
\newcommand{\rb}{{\bf r}}

\centering \huge Machine Learning Frontier Orbital Energies of Nanodiamonds \\ ~\\
\large Thorren Kirschbaum\textsuperscript{1,2}, Börries von Seggern\textsuperscript{1,3}, Joachim Dzubiella\textsuperscript{1,4}, Annika Bande\textsuperscript{1$\ast$}, Frank Noé\textsuperscript{5,2,6,7$\ast$} \\ ~\\
\raggedright \normalsize \textsuperscript{1}Helmholtz-Zentrum Berlin für Materialien und Energie GmbH, Hahn-Meitner-Platz 1, 14109 Berlin, Germany\\
\textsuperscript{2}Department of Mathematics and Computer Science, Freie Universität Berlin, Arnimallee 12, 14195 Berlin, Germany\\
\textsuperscript{3}Department of Biology, Chemistry and Pharmacy, Freie Universität Berlin, Arnimallee 22, 14195 Berlin, Germany\\
\textsuperscript{4}Institute of Physics, Albert-Ludwigs-Universität Freiburg, Hermann-Herder-Straße 3, 79104 Freiburg im Breisgau, Germany\\
\textsuperscript{5}Microsoft Research AI4Science, Karl-Liebknecht Str. 32, 10178 Berlin, Germany\\
\textsuperscript{6}Department of Physics, Freie Universität Berlin, Arnimallee 12, 14195 Berlin, Germany\\
\textsuperscript{7}Department of Chemistry, Rice University, 6100 Main St, Houston, TX 77005, United States\\
 ~\\
\textsuperscript{$\ast$}Corresponding Author: annika.bande@helmholtz-berlin.de\\
\textsuperscript{$\ast$}Corresponding Author: franknoe@microsoft.com

\clearpage

\begin{abstract}

\begin{justify}

Nanodiamonds have a wide range of applications including catalysis, sensing, tribology and biomedicine. To leverage nanodiamond design via machine learning, we introduce the new dataset ND5k, consisting of 5,089 diamondoid and nanodiamond structures and their frontier orbital energies. ND5k structures are optimized via tight-binding density functional theory (DFTB) and their frontier orbital energies are computed using density functional theory (DFT) with the PBE0 hybrid functional. 
We also compare recent machine learning models for predicting frontier orbital energies for similar structures as they have been trained on (interpolation on ND5k), and we test their abilities to extrapolate predictions to larger structures. For both the interpolation and extrapolation task, we find best performance using the equivariant graph neural network PaiNN. The second best results are achieved with a message passing neural network using a tailored set of atomic descriptors proposed here.

\end{justify}

\end{abstract}

\clearpage

\section{Introduction}\label{SecIntro}

\begin{justify}

Research on nanomaterials has experienced a rapid upswing in recent decades.\cite{roduner_size_2006, baig_nanomaterials_2021} Among the different material classes, nanodiamonds stand out due to their unusually high stability, biocompatibility, and notable electronic properties, leading to applications in catalysis, sensing, tribology, biomedicine, and many more.\cite{mochalin_properties_2012, nunn_nanodiamond_2017} In the field of photocatalysis, nanodiamonds are used as a source of electrons, enabling high-energy reduction reactions in the liquid phase, including the reductions of N$_2$ to ammonia, of CO$_{2}$ to CO, and of water into its elements.\cite{zhu_photo-illuminated_2013, zhang_selective_2014, hamers_photoemission_2014, zhang_photocatalytic_2017}

A large body of scientific literature has been dedicated to the investigation of specific nanodiamond properties, both from experimental and theoretical perspectives.\cite{lai_surface_2012, petit_valence_2015, petit_unusual_2017, choudhury_combining_2018, feigl_classifying_2019, kirschbaum_effects_2022} At the same time, the design of diamond-based materials remains a hot topic for several applications.\cite{teunissen_tuning_2017, zhang_hybrid_2018, wang_nanodiamonds_2019, liu_nanodiamond-enabled_2020} The most prominent approaches for “tuning” the nanodiamonds’ properties are size modification,\cite{chang_quantum_1999, willey_molecular_2005, stehlik_size_2015} surface functionalization\cite{ferro_physicochemical_2003, wang_comparison_2009, brown_controlling_2014, larsson_effect_2018} and the incorporation of dopant atoms.\cite{pinault_n-type_2007, williams_growth_2008, knittel_nanostructured_2020} Further design choices include the degree of surface graphitization on the nanodiamonds\cite{pichot_efficient_2008, mikheev_low-power_2020} and the use of composite materials to complement and enhance their properties.\cite{behler_nanodiamond-polymer_2009, bandy_photocatalytic_2016, etemadi_performance_2017, petit-dominguez_synergistic_2018} The wide range of design possibilities renders a large array of possible nanodiamond configurations of which only a few can be tested in dedicated individual studies, ultimately resulting in slow design processes. 

High-throughput screening has recently emerged as a pathway for rapidly identifying candidate materials for a given application.\cite{dral_quantum_2020} In this approach, a wide range of materials is tested for a certain set of properties using an inexpensive theoretical or experimental method. Subsequently, the most promising materials can further be researched in more elaborate studies. Theory-based algorithms that can perform such inexpensive predictions either come from the field of quantum chemistry (e.g., tight-binding density functional theory, DFTB),\cite{foulkes1989tight} or from the field of machine learning (ML).\cite{dral_quantum_2020} In supervised ML, generally, a suitable algorithm learns to perform a desired mapping, e.g., that from a chemical structure to its HOMO-LUMO gap, given a dataset of examples (training set). The ML workflow consists of obtaining this training set, choosing a suitable ML algorithm, training the algorithm on the training set and finally testing its performance on unseen data. The trained ML algorithm is then able to perform the mapping much faster than the corresponding quantum chemical approaches. 

Machine-learning frontier orbital energies of molecules and materials has been the subject of previous investigations. A range of studies focused on learning frontier orbital energies and related properties for optimizing solar photovoltaic materials.\cite{li_machine-learning_2018, lee_insights_2019, padula_combining_2019, lee_machine_2020, meftahi_machine_2020, lee_identifying_2022, zhang_high-efficiency_2022, storm_machine_2022} Further studies explored the use of different molecular descriptors,\cite{pereira_machine_2017, li_machine-learning_2018, mchang_hammett_2019, olsthoorn_band_2019, rahaman_deep_2020, woon_relating_2021, ye_assessment_2022} investigated structure-property relationships,\cite{woon_relating_2021, zhang_high-efficiency_2022} compared different DFT functionals,\cite{duan_machine_2021} and presented improved learning strategies.\cite{mazouin_selected_2022}

For ML in chemistry, mainly two types of learning algorithms have become popular over the last years, atomic fingerprinting techniques and graph neural networks (GNNs).\cite{nigam_unified_2022} Both can operate on molecules of arbitrary size and are usually designed to inherently obey the invariances and equivariances of physical properties with respect to global rotation and translation of the molecular system, as well as permutation of indistinguishable atoms.

In atomic fingerprinting techniques, atom-wise symmetry functions are calculated for each atom of the system.\cite{behler_generalized_2007, bartok_representing_2013, parsaeifard_assessment_2021} These functions describe the local chemical environment of an atom in a rotation- and translation-invariant manner. Two of the most popular algorithms are the atom-centered symmetry functions as introduced by Behler and Parinello,\cite{behler_generalized_2007} and the smooth overlap of atomic positions (SOAP) descriptor.\cite{bartok_representing_2013} Fingerprinting approaches are often used with kernel ridge regression (KRR) for the prediction of atomic or molecular properties.\cite{musil_physics-inspired_2021}

GNNs for molecules are neural networks which, instead of taking dedicated feature functions as inputs, directly process atom positions and types and learn a representation of suitable features internally. They represent molecular structures as graphs, with nodes representing the atoms and edges representing their interactions (either bonded or nearby nonbonded).\cite{kipf2016, gilmer_neural_2017, schutt_schnet_2017, schutt_quantum-chemical_2017, thomas2018, schutt_equivariant_2021, miller_relevance_2020, batzner_e3-equivariant_2022, batatia2022, geiger2022} Each graph node (and sometimes also each edge) is assigned a feature vector that carries information on the atom (or bond) it represents, often starting with a one-hot encoding of the atom (bond) type. During training, these features are updated by graph convolutions that share parameters in such a way that permutation invariance or equivariance is maintained. The convolution kernels typically depend upon distances or angular information between atoms, such that translational and rotational invariance or equivariance is also maintained. The prediction of the desired molecular property is usually made in a final readout step, e.g. by summing or averaging the feature vectors after the last convolution layer, by processing them with a densely connected network, or applying more elaborate aggregation functions.\cite{gilmer_neural_2017}

In order to balance bias and variance, simpler models such as KRR typically have lower prediction error for little data. On the other hand, more expressive models such as deep NNs tend to be superior in the large-data limit, and also exhibit better computational cost scaling to large amounts of data and large numbers of chemical species. Accordingly, GNNs have recently achieved top-level performances on the large standard chemistry ML databases,\cite{miller_relevance_2020, schutt_equivariant_2021, batzner_e3-equivariant_2022} however, their application to new problems is often limited by the amount of data available for the learning task at hand. A few large chemistry datasets containing DFT-computed properties of small molecules are freely available.\cite{blum_970_2009, ruddigkeit_enumeration_2012, montavon_machine_2013, ramakrishnan_quantum_2014, draxl_nomad_2019, stuke_atomic_2020} However, systematically crafted datasets of nanomaterials containing molecular data obtained from a high level of theory usually consist of only a few hundred data points, rendering them insufficient for their use in large-scale design approaches.\cite{fernandez_machine_2017, sun_machine_2017, barnard_predicting_2019, barnard_does_2019, furxhi_machine_2019, jparker_machine_2020, feigl_classifying_2019, weber_theoretical_2019, daly_learning_2020} Such databases are highly desirable for the design of nanodiamonds as, i.e., photocatalysts,\cite{zhu_photo-illuminated_2013, zhang_selective_2014, hamers_photoemission_2014, zhang_photocatalytic_2017} electronic devices\cite{luong_boron-doped_2009, denisenko_surface_2010, acosta_nitrogen-vacancy_2013} and energy materials.\cite{wang_nanodiamonds_2019}

In this paper, we introduce the ND5k dataset, which consists of 5,089 diamondoid and nanodiamond structures optimized via DFTB. In addition, we report their frontier orbital energies, which are key properties for the aforementioned applications, computed at DFT PBE0-D3/SVP level of theory. Diamondoids are molecular-scale nanodiamonds, the smallest being adamantane (C$_{10}$H$_{16}$), while the largest structure of the dataset is about 1.2 nm in size and contains 189 heavy (non-H) atoms. From the chemistry ML perspective, this dataset provides a natural extension to existing databases of organic molecules (QM7b,\cite{blum_970_2009, montavon_machine_2013} QM9,\cite{ruddigkeit_enumeration_2012, ramakrishnan_quantum_2014} OE62,\cite{stuke_atomic_2020} etc.) towards larger and more complex carbon-based structures. In the present study, we analyze general trends in the ND5k dataset and extract design choices for nanodiamonds in photocatalysis. Furthermore, we test latest ML approaches for use with this dataset and discuss their performances as well as their abilities to extrapolate their predictions towards even larger nanodiamonds. In particular, we compare the SOAP KRR approach\cite{bartok_representing_2013, musil_physics-inspired_2021} with three GNNs, an edge-conditioned GNN with set2set pooling (enn-s2s),\cite{gilmer_neural_2017} as well as SchNet\cite{schutt_schnet_2017, schutt_quantum-chemical_2017} and the equivariant PaiNN.\cite{schutt_equivariant_2021} In this context, we propose an extension of GNNs to take advantage of both molecular fingerprints and GNN learning architectures, which we apply to the enn-s2s architecture.

\end{justify}

\clearpage

\section{Methods}\label{SecMethods}

\subsection{Dataset Computations}

\begin{justify}

Diamondoid and nanodiamond structures were generated in a semi-automated fashion. Structure optimization was performed via DFTB using Grimme's extended tight binding (xtb) software,\cite{bannwarth_extended_2021} version 6.2.3, the GFN2-xTB method\cite{bannwarth_gfn2_2019} and default settings for geometry optimization. The validity of the DFTB optimization is discussed in more detail in the supporting information. If steric hindrance on the surface of the highly decorated nanodiamonds was too strong, surface groups were removed and replaced with hydrogen (which was especially common for amine- and hydroxyl-terminated nanodiamonds). For the case of N dopants in the nanodiamond lattice, one neighboring C atom was removed to form a nitrogen-vacancy structure, if (and only if) the structure without a vacancy was not stable. Sanity checks were performed to ensure that chemically reasonable structures were generated. Subsequently, single point calculations were performed within the ORCA 5 software package,\cite{neese_orca_2020} using the PBE0 hybrid functional,\cite{adamo_toward_1999, perdew_rationale_1996} Ahlrich’s def2-SVP basis set\cite{schafer_fully_1992, weigend_balanced_2005} and Grimme’s third-order atom-pairwise dispersion correction with Becke-Jones dampening (D3BJ).\cite{grimme_effect_2011} The RIJCOSX approximation and an appropriate auxiliary basis set were used to speed up integral calculations, as implemented in the ORCA default settings.\cite{neese_efficient_2009, weigend_accurate_2006} We used the ORCA DEFGRID1 integration grid option for fast integral calculations, the deviations to the more accurate DEFGRID2 default option are in the order of 0.01 eV for the frontier orbital energies.

\end{justify}

\subsection{Machine Learning Models}

\begin{justify}

In supervised ML, as a rule of thumb, kernel methods are to be preferred when less than ~ 10$^3$–10$^4$ training points are available, and NN-based approaches otherwise.\cite{unke_machine_2021}. Our present dataset has an intermediate size, and will therefore compare both types of methods for machine learning the frontier orbital energies of the ND5k structures.

The first ML setup uses the SOAP atomic fingerprinting technique in conjunction with KRR. Following Parsaeifard \textit{et al.},\cite{parsaeifard_assessment_2021} in the SOAP approach, a Gaussian of width $\sigma$ is centered on each atom within the cutoff distance of a central atom $k$ at position \textbf{r$_k$}. The resulting density of atoms is multiplied with a cutoff function $f_{cut}$ which smoothly approaches zero at a cutoff radius over a characteristic width r$_{\delta}$,

\begin{equation}
\rho ^k(\textbf{r}) = \sum\nolimits _i \mathrm{exp} \left( - \frac {\left(\textbf{r}-\textbf{r}_{ki}\right)^2} {2 \sigma ^2} \right) \times f_{cut}\left( \left| \textbf{r}-\textbf{r}_{ki} \right| \right).
\end{equation}

This density is then expanded in terms of orthogonal radial functions $g_n(r)$ and spherical harmonics $Y_{lm}(\theta, \phi)$ as 
\begin{equation}
\rho^k(\textbf{r}) = \sum\nolimits _{nlm} c^k_{nlm} g_n(r) Y_{lm} (\theta, \phi),
\end{equation}
where
$
c^k_{nlm} = \langle g_n Y_{lm} | \rho^k \rangle
$. Finally,
\begin{equation}
p^k_{nn'l} = \sqrt{\frac {8\pi^2} {2l+1} } \sum\nolimits_m c^k_{nlm} \left( c^k_{n'lm} \right)^{*}
\end{equation}
are obtained as rotationally invariant scalar descriptors of the central atom $k$'s environment. The SOAP fingerprint vector \textbf{F}$^k$ then contains all $p^k_{nn'l}$ with $n$, $n'$ $\leq$ $n_{max}$ and $l$ $\leq$ $l_{max}$.\cite{parsaeifard_assessment_2021}

Second, we use the enn-s2s GNN as proposed by Gilmer \textit{et al.}\cite{gilmer_neural_2017} Within this approach, the undirected molecular graph $G$ is initiated with atomic feature vectors $x_v = h_v^0$ of size $d$ at the graph nodes, containing a set of chemical properties of the respective element (e.g., one-hot atom type, atomic number, hybridization state, etc.), and bond features $e_{vw}$ at the graph edges (containing, e.g., one-hot bond type and bond distance). The hidden states $h_v^0$ of the nodes then get updated $T$ times during the message passing phase, which is defined in terms of message functions $M_t$ that generate messages $m_v^{t+1}$, and node update functions $U_t$:
\begin{equation}
m_v^{t+1} = \sum\nolimits_{w\in N(v)} M_t (h_v^t, h_w^t, e_{vw})
\end{equation}
with
$h_v^{t+1} = U_t (h_v^t, m_v^{t+1}) $,
where $w\in N(v)$ denotes the neighbors $w$ of node $v$ in graph $G$. Here, the message function has the form 
$M(h_v,h_w,e_{vw}) = A(e_{vw})h_w$,
where $A(e_{vw})$ is a dense NN that maps the edge vector $e_{vw}$ to a $d \times d$ matrix, and a gated recurrent unit (GRU)\cite{cho2014properties} is used for the update function. After the message passing, the graph information is accumulated using a set2set operator,\cite{vinyals2015order} whose output is then passed to a dense NN that computes the final prediction.\cite{gilmer_neural_2017} 

In the recent GNN developments, one of the main improvements was the increased ability of the networks to harness structural information for their predictions. While earlier variants such as the enn-s2s proposed by Gilmer \textit{et al.}\cite{gilmer_neural_2017} and SchNet\cite{schutt_schnet_2018, schutt_schnetpack_2019} were only able to use atomic distances for their predictions, more recent approaches, such as DimeNet,\cite{klicpera2020directional, klicpera2020fast} L1Net\cite{miller_relevance_2020} and PaiNN,\cite{schutt_equivariant_2021} are able to efficiently make use of angular information as well. Following the same idea, we propose to replace the atom-type-specific node embeddings of GNNs with atom-centered descriptors of the atoms' local environments. Furthermore, these descriptors may be condensed by applying dimensionality reduction techniques, such as principle component analysis (PCA).\cite{abdi_principal_2010, casier_using_2020, parsaeifard_assessment_2021, darby_compressing_2022} In this study, we implement two variants of the enn-s2s: The first one uses SOAP node embeddings (SOAP-enn-s2s) for the graph setup, and the second one uses the PCA-reduced SOAP vectors (SOAP-PCA-enn-s2s) instead of generic atomic descriptors as initial node embeddings of the graphs. For the latter, the 1548-dimensional SOAP vectors are reduced to 133-dimensional vectors via PCA, thereby retaining 99.99 \% of their variance.

The other type of GNN which we apply, SchNet, takes an alternative approach.\cite{schutt_schnet_2018, schutt_schnetpack_2019} Here, the node representations are initialized randomly for each element type and subsequently adapted (learned) during training. During the message passing, continuous-filter convolutions are used to incorporate the influence of neighboring atoms. After passing $N$ such interaction layers, a prediction block is used to obtain the final result. For the prediction of molecular properties, a sum-over-nodes operation (sum pooling) is used to obtain extensive properties and an average pooling to obtain intensive properties.\cite{schutt_schnet_2018, schutt_schnetpack_2019}

Recently, the polarizable atom interaction NN (PaiNN) was proposed as an extension of SchNet.\cite{schutt_equivariant_2021} The PaiNN architecture uses equivariant representations over angular features to enable the incorporation of angular information into the learning process. Thus, the network receives additional structural data while retaining the ability to make rotation-invariant predictions by design.\cite{schutt_equivariant_2021}

\end{justify}

\subsection{Machine Learning Details}

\begin{justify}

Machine learning was performed within Python, using the libraries \textit{librascal}~\cite{musil2018librascal} and \textit{scikit-learn}~\cite{scikit-learn} for KRR with SOAP, and the \textit{pytorch},\cite{DBLP:journals/corr/abs-1912-01703} \textit{pytorch-geometric}\cite{Fey/Lenssen/2019} and \textit{schnetpack}\cite{schutt_schnetpack_2019} libraries for deep learning with GNNs. The SOAP vectors used as inputs for the SOAP node embedding (SOAP) enn-s2s approach were computed using the \textit{Dscribe} package\cite{dscribe} in conjuction with the atomic simulation environment package (\textit{ase}) as described by Larsen \textit{el al}.\cite{Hjorth_Larsen_2017} The dataset was split into 66 \% train and 17 \% test subsets (KRR), or 66 \% train, 17 \% validation and 17 \% test subsets (NN training), using a stratified split. The six low LUMO energy outliers at E $< -$6 eV were removed for learning LUMO energies. For NN training, early stopping was applied. The enn-s2s architecture is based on the continuous kernel-based convolutional operator NNConv\cite{gilmer_neural_2017, Fey/Lenssen/2019} and inspired by the work of Ramahan \textit{et al.} who used a similar architecture for learning frontier orbital energies and total energies of large organic molecules.\cite{rahaman_deep_2020} For PaiNN, we used the implementation provided alongside the original publication.\cite{schutt_equivariant_2021} Further details on the ML setup are available in the supporting information.

\end{justify}

\clearpage

\section{Results}\label{SecResults}

\subsection{The ND5k Dataset}

\begin{justify}

The ND5k dataset is publicly available as described in the data availability section \ref{sec:data_availability}.
Structures in ND5k were generated systematically, as indicated in the flowchart in figure \ref{fig:flowchart}. We selected 17 structures of undoped, H-terminated diamondoids and small nanodiamonds as base structures for the dataset. Diamondoids are nanodiamonds of molecular size, of which the smallest, adamantane (C$_{10}$H$_{16}$), consists of only one cage of tetrahedrally coordinated carbon atoms. The larger diamondoids can be built from this structure by subsequentially adding further diamond cages to this structure. The largest base structure in the dataset is a nanodiamond with sum formula C$_{109}$H$_{80}$ and a diameter of ca. 1.2 nm. An overview of all base structures is provided in the supporting information, figure S1. 

To create a more diverse set of structures, each base nanodiamond was covered with each of the surface moieties H, F, OH or NH$_2$ and added to the dataset, as these are common surface decorations in nanodiamond synthesis.\cite{liu_functionalization_2004, wang_comparison_2009, panich_structure_2010, shenderova_hydroxylated_2011, zhu_amino-terminated_2016}

In a third step, each of these structures was doped with one or two dopant atoms, including B, N, Si and P, and the resulting structures were added to the dataset. While B doping of diamond materials can readily be achieved experimentally,\cite{denisenko_surface_2010, choudhury_combining_2018, knittel_nanostructured_2020} n-type doping remains a challenge so far.\cite{kajihara_nitrogen_1991, pinault_n-type_2007} Moreover, doping with Si\cite{cui_si-doped_2013, wang_influences_2020, wei_effect_2020} and P\cite{grotjohn_heavy_2014, kato_heavily_2016, alfieri_phosphorus-related_2018} is especially difficult, because the small lattice parameters and its high stiffness hinder the incorporation of large dopants into the diamond lattice. Still, we include these dopants in the dataset to cover a wider range of structures and to acknowledge the fact that in principle, doping with these atoms is possible. 

For each combination of base structure, surface covering and (co-)doping, three different structures with varying positions of the dopant atoms are included in the dataset. This combinatorial approach results in a total of 2,908 structures. Finally, to obtain not only fully, but also partially functionalized nanodiamond structures, for the F-, OH- and NH$_2$-terminated structures, a random number of surface moieties was removed and replaced by hydrogen and the new structure added to the dataset. The final dataset then contains 5,089 structures in total. Note that for each singly doped adamantane structure only two distinct dopant positions are possible and therefore, only two structures could be inserted instead of three. 

\begin{figure}[h]
 \raggedright
 \includegraphics[width = 0.78\textwidth]{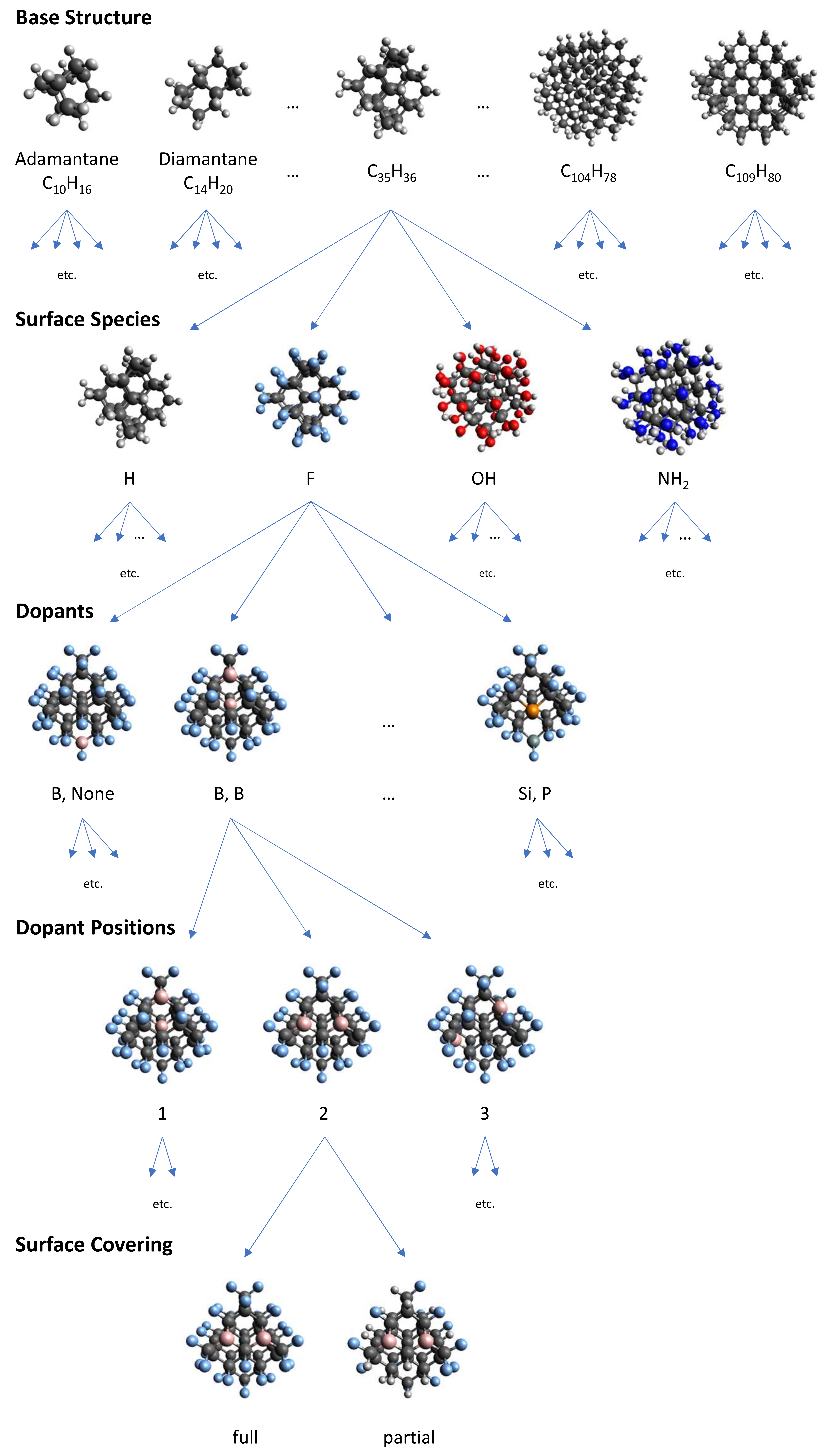}
    \caption{Flowchart illustrating the composition of the ND5k structures. The example shows the nanodiamonds with C$_{35}$H$_{36}$ base structure, F surface covering, two B dopants, and full or partial F-termination. Color: H (grey), B (rose), C (black), N (dark blue), O (red), F (light blue), Si (green,) P (orange).}
    \label{fig:flowchart}
\end{figure} 

All ND5k structures were optimized using DFTB and their single point properties calculated within standard DFT using the PBE0 hybrid functional (see Methods section). In this work, we focus on the energies of the nanodiamonds' frontier orbitals that were obtained from the calculations, i.e., the highest occupied molecular orbital (HOMO) and the lowest unoccupied molecular orbital (LUMO), as these are most relevant for applications such as electronic devices and photocatalysis. 

Figure \ref{fig:dist} shows a histogram of the ND5k HOMO and LUMO energies that were obtained at PBE0 level of theory. The HOMO energies range from $-$10.6 to $+$0.3 eV, and most structures have a HOMO energy between $-$7 and $-$6 eV. One set of structures is separated from the main distribution and locates at lower energy values E $< -$ 9 eV, representing the highly fluorinated diamondoids and nanodiamonds. Another small set of structures has higher energy values E $> -$2 eV which are mostly H-terminated, P-doped larger nanodiamonds. The LUMO energies range from $-$8.6 to $+$1.6 eV, and most structures have a LUMO energy around $+$1 eV. This is consistent with the fact that nanodiamonds, especially when H- or NH$_2$-terminated, are known to have a negative electron affinity.\cite{ristein_surface_2006} The LUMO distribution has a longer tail towards more negative energies, and six outliers can be identified at E $< -$6 eV. These six structures are larger nanodiamonds with full or partial F-termination and two B dopants incorporated into the nanodiamond lattice.

\clearpage

\begin{figure}[t]
 \raggedright
 \includegraphics[width = 0.8\textwidth]{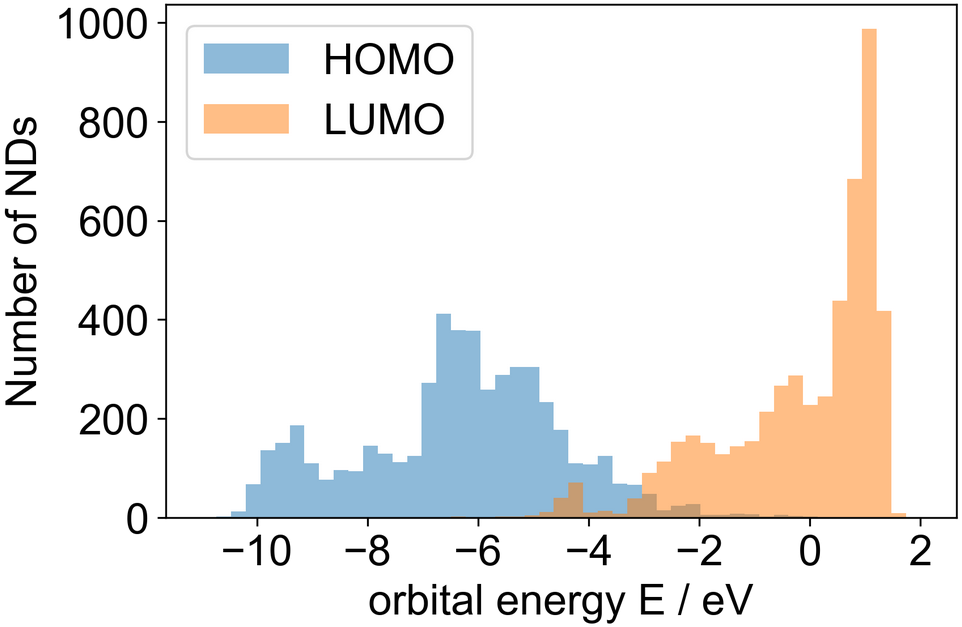}
    \caption{HOMO (blue) and LUMO (orange) orbital energy distributions of the ND5k dataset.}
    \label{fig:dist}
\end{figure}

Very specific nanodiamond properties are desired for each individual application, according to the specific use case. For example, in photocatalysis, nanodiamonds with negative electron affinity are needed to enable facile emission of electrons from the nanodiamond upon light excitation.\cite{zhu_photo-illuminated_2013, zhang_selective_2014, hamers_photoemission_2014, zhang_photocatalytic_2017, buchner_early_2021} The electrons that are generated in this way are potent reduction agents which can initiate high-energy reduction reactions, e.g., the reduction of N$_2$ to ammonia.\cite{zhu_photo-illuminated_2013, christianson_mechanism_2014} The reaction is usually performed in aqueous environments: Here, one electron is excited into the nanodiamonds' conduction band and then moves into the unoccupied states of the water bulk. For our purposes, the nanodiamond's conduction band minimum energy can be approximated as the LUMO energy of the molecular nanodiamond. Similarly, the adjacent water can be regarded as an amorphous semiconductor with a conduction band minimum at around $-$1 eV into which the electron is transferred.\cite{bandy_photocatalytic_2016, ambrosio_absolute_2018} To enable the ejection of an electron, the LUMO (or conduction band minimum) of the nanodiamond needs to be energetically above the water’s conduction band minimum. Furthermore, it is desirable to drive the reaction using sunlight, meaning that the nanodiamond’s HOMO-LUMO gap should be at around 2.6 eV or lower. 

According to these criteria, we filtered a list of 45 candidate structures from our dataset, with LUMO energies $\ge -$1 eV and gap energies $\leq$ 2.6 eV, which are presented in table \ref{tab:NDs_photocat}. The proposed structures are mostly larger nanodiamonds with either H or NH$_2$ surface decoration and mostly doped with P. Owing to the quantum confinement effect, larger structures generally have smaller optical gaps, and are thus better suited for this approach. The H and NH$_2$ surface coverings facilitate adequately high LUMO energies and concomitant negative electron affinities, as demonstrated in various experimental studies.\cite{zhu_photo-illuminated_2013, christianson_mechanism_2014, zhang_selective_2014, hamers_photoemission_2014, bandy_photocatalytic_2016, zhang_photocatalytic_2017} Finally, the n-type doping introduces new occupied states into the nanodiamond gap, from which electrons can facilely be excited into the unoccupied states. Here, phosphorous seems to be preferable over nitrogen doping, because the excess electrons introduced by the P dopants are located at relatively higher energy levels. 
In the supporting information, we plot the structures and frontier orbital shapes of ten P-(co-)doped nanodiamonds taken from table \ref{tab:NDs_photocat}. For all structures, the HOMO is located at the respective P dopant atom. The LUMO is located at the surface of the nanodiamond in 8 of 10 structures, and located at the second dopant (boron) in two structures. For one representative structure (C$_{48}$H$_{48}$ base structure, one single phosphorous dopant, hydrogen surface termination, index 3001), the HOMO and LUMO orbitals are depicted in figure \ref{fig:ND3001}. The HOMO is well localized at the P dopant, and the diffuse LUMO is spread out over the surface of the whole structure. It is well known that spherical nanodiamonds,\cite{han_surface-bound_2017} diamondoids, and clusters of diamondoids\cite{kirschbaum_effects_2022} have uniquely shaped unoccupied orbitals: For these structures, the LUMOs resemble atomic s-orbitals, and next highest unoccupied orbitals are shaped like atomic p-, d-, and f-orbitals. The same type of unoccupied orbitals are present in the nanodiamond shown here, with slight distortions due to the low symmetry of the structure. Thus, the phosphorous doping of the nanodiamond introduces a new occupied electronic level, while the overall electronic structure is retained. Accordingly, hydrogen- or amine-covered, phosphorous-doped nanodiamonds are promising candidate materials to perform sunlight-driven electron emission in the aqueous phase. This finding can be regarded as a qualitative design suggestion for future research in this field.

Different nanodiamond applications may rely on other values of the particles' frontier orbital alignments. In the supporting information, we summarize the ND5k structures which have the highest and lowest HOMO, LUMO and gap energies, respectively, and shortly discuss the general trends.

\clearpage

\begin{table}[h]
\label{tab:NDs_photocat}

\caption{Overview of the 45 ND5k structures with LUMO energies $\ge$ $-$1 eV and gap energies $\leq$ 2.6 eV for application as photocatalysts: ND5k index, base structure (ND), surface species, dopants (D1, D2), HOMO, LUMO and gap energies (all in eV). The base structures abbreviated as  123tet, 1212pent and 12312hex are the diamondoids [123]tetramantane, [1212]pentamantane and [12312]hexamantane, respectively (see also the overview of base structures in the SI).}

\medskip
\footnotesize
\begin{tabularx}\columnwidth{XXXXXXXX}

Index & ND & Surface & D1 & D2 & E(HOMO) & E(LUMO) & E(gap) \\
\hline
2416     & 12312hex & H     & P  & P  & $-$1.649  & 0.942   & 2.591 \\
3785     & C$_{68}$H$_{64}$   & OH/H  & P  & P  & $-$2.011 & 0.574   & 2.585 \\
3892     & C$_{74}$H$_{64}$   & H     & B  & N  & $-$1.939 & 0.628  & 2.567 \\
3641     & C$_{68}$H$_{64}$   & NH$_2$   & B  & N  & $-$3.219 & $-$0.652 & 2.567 \\
3001     & C$_{48}$H$_{48}$   & H     & P  &   & $-$1.435 & 1.129  & 2.564 \\
2701     & C$_{35}$H$_{36}$   & H     & P  &   & $-$1.430 & 1.073  & 2.503 \\
3614     & C$_{68}$H$_{64}$   & H     & N  & P  & $-$1.324 & 1.174  & 2.498 \\
4693     & C$_{104}$H$_{78}$  & OH/H  & P  & Si & $-$1.688 & 0.767  & 2.456 \\
1485     & 123tet   & H     & B  & N  & $-$1.870 & 0.584  & 2.454 \\
3992     & C$_{74}$H$_{64}$   & NH$_2$   & N  & Si & $-$3.160 & $-$0.752 & 2.408 \\
3923     & C$_{74}$H$_{64}$   & H     & P  & Si & $-$1.456 & 0.909  & 2.365 \\
4874     & C$_{109}$H$_{80}$  & NH$_2$/H & P  &   & $-$1.654 & 0.708  & 2.363 \\
3998     & C$_{74}$H$_{64}$   & NH$_2$   & P  & P  & $-$1.248 & 1.054  & 2.301 \\
4001     & C$_{74}$H$_{64}$   & NH$_2$/H & P  & P  & $-$1.332 & 0.952  & 2.284 \\
2878     & C$_{35}$H$_{36}$   & OH    & P  & P  & $-$2.052 & 0.223  & 2.274 \\
1789     & 1212pent & H     & B  & P  & $-$2.874 & $-$0.617 & 2.257 \\
4903     & C$_{109}$H$_{80}$  & NH$_2$/H & P  & P  & $-$1.399 & 0.810  & 2.209 \\
3922     & C$_{74}$H$_{64}$   & H     & P  & Si & $-$1.424 & 0.693  & 2.117 \\
4524     & C$_{104}$H$_{78}$  & H     & P  & Si & $-$0.930   & 1.184  & 2.114 \\
4600     & C$_{104}$H$_{78}$  & NH$_2$   & P  & P  & $-$1.069 & 1.028  & 2.097 \\
2743     & C$_{35}$H$_{36}$   & NH$_2$   & B  & P  & $-$2.543 & $-$0.455 & 2.088 \\
3012     & C$_{48}$H$_{48}$   & H     & N  & P  & $-$1.233  & 0.849  & 2.082 \\
3634     & C$_{68}$H$_{64}$   & NH$_2$   & B  &   & $-$2.417 & $-$0.357 & 2.060  \\
2746     & C$_{35}$H$_{36}$   & NH$_2$/H & B  & P  & $-$2.424 & $-$0.391 & 2.033 \\
4598     & C$_{104}$H$_{78}$  & NH$_2$   & P  & P  & $-$0.968 & 0.962  & 1.930  \\
4220     & C$_{88}$H$_{80}$   & H     & P  & P  & $-$0.990 & 0.906  & 1.896 \\
4505     & C$_{104}$H$_{78}$  & H     & N  &   & $-$0.660 & 1.161  & 1.822 \\
3620     & C$_{68}$H$_{64}$   & H     & P  & P  & $-$0.943 & 0.870  & 1.813 \\
3315     & C$_{53}$H$_{48}$   & H     & N  & Si & $-$1.101 & 0.623  & 1.725 \\
4809     & C$_{109}$H$_{80}$  & H     & P  &   & $-$0.540   & 1.104  & 1.644 \\
4601     & C$_{104}$H$_{78}$  & NH$_2$/H & P  & P  & $-$0.740 & 0.884  & 1.624 \\
5074     & C$_{109}$H$_{80}$  & F/H   & P  & P  & $-$2.155  & $-$0.590 & 1.565 \\
4602     & C$_{104}$H$_{78}$  & NH$_2$/H & P  & P  & $-$0.540 & 1.01    & 1.550  \\
3314     & C$_{53}$H$_{48}$   & H     & N  & Si & $-$0.957 & 0.557  & 1.514 \\
4599     & C$_{104}$H$_{78}$  & NH$_2$   & P  & P  & $-$0.277 & 1.160  & 1.437 \\
3397     & C$_{53}$H$_{48}$   & NH$_2$/H & P  & P  & $-$0.514 & 0.909  & 1.423 \\
4825     & C$_{109}$H$_{80}$  & H     & P  & Si & $-$0.565 & 0.750  & 1.316 \\
4521     & C$_{104}$H$_{78}$  & H     & P  & P  & $-$0.383  & 0.838  & 1.221 \\
3394     & C$_{53}$H$_{48}$   & NH$_2$   & P  & P  & $-$0.592 & 0.621  & 1.214 \\
4823     & C$_{109}$H$_{80}$  & H     & P  & P  & $-$0.381 & 0.799  & 1.180  \\
3016     & C$_{48}$H$_{48}$   & H     & P  & P  & $-$0.500 & 0.675  & 1.174 \\
3647     & C$_{68}$H$_{64}$   & NH$_2$   & B  & P  & $-$1.580 & $-$0.518 & 1.062 \\
2711     & C$_{35}$H$_{36}$   & H     & N  & P  & $-$0.380 & 0.412  & 0.792 \\
2709     & C$_{35}$H$_{36}$   & H     & N  & P  & $-$0.097 & 0.522   & 0.619 \\
3921     & C$_{74}$H$_{64}$   & H     & P  & P  & 0.033  & 0.557  & 0.525 \\
\end{tabularx}
\normalsize
\end{table}

\begin{figure}[H]
 \raggedright
 \includegraphics[width = 0.35\textwidth]{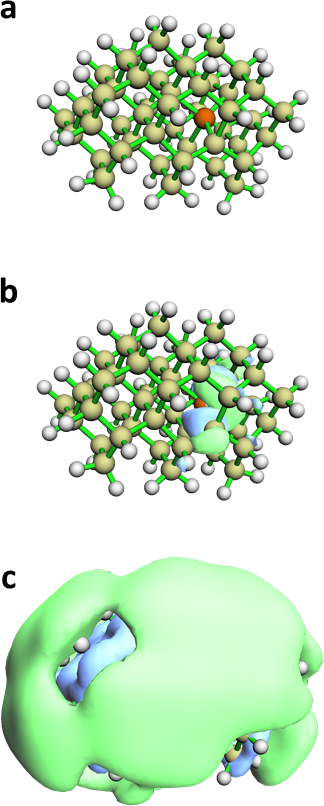}
    \caption{a) Structure, b) HOMO contour plot, c) LUMO contour plot of the nanodiamond with ND5k index 3001 (P-doped, H-terminated, C$_{48}$H$_{48}$ base structure). Color: H (grey), C (yellow), P (orange). The well localized HOMO is plotted with isovalue $\pm$ 0.05, the diffuse LUMO with isovalue $\pm$ 0.01.}
    \label{fig:ND3001}
\end{figure} 

\clearpage

\end{justify}

\subsection{Predicting Frontier Orbital Energies with Machine Learning}

\begin{justify}

We use the six ML algorithms introduced before (SOAP KRR, enn-s2s, SOAP-enn-s2s, SOAP-PCA-enn-s2s, SchNet, PaiNN) to predict the frontier orbital energies of the structures in the ND5k dataset. For learning LUMO energies, we remove the six low-energy-LUMO outliers (E(LUMO) $<$ $-$6 eV) from the dataset. In all cases, hyperparameters were optimized by an iterative grid search, and the final values are listed in the supporting information. In addition to the ND5k learning task, we compiled a test set of 24 nanodiamonds with larger base structures than available in the ND5k dataset and random surface covering and doping patterns, in accordance with the ND5k setup. This ND5k large test set, referred to as ND5k-lt, was used to evaluate the extrapolation capabilities of the ML algorithms towards larger structures. For PaiNN, we use both sum pooling and average pooling after the message passing phase, while SchNet use average pooling only and the enn-s2s variants use set2set (learnable) pooling only.

The mean absolute errors (MAE) obtained by different ML methods are summarized in table \ref{tab:ML}, learning curves are plotted in figure \ref{fig:LCs}. Results for all algorithms are averaged over 6 random splits. An overview of the models' hyperparameters and a more detailed description of the enn-s2s architecture are given in the supporting information.

\begin{table}[h]
\caption{Mean absolute errors (MAE, in eV) for ND5k HOMO and LUMO energy predictions, and errors for predicting HOMO and LUMO energies of 24 larger nanodiamonds (ND5k-lt) after training on ND5k. Uncertainties are measured in root mean squared error (RMSE) (in parentheses). Best MAE results are bold.}
\medskip
\begin{tabular}{lllll}
ML Method & \begin{tabular}[c]{@{}l@{}}ND5k \\ HOMO Energy\end{tabular} & \begin{tabular}[c]{@{}l@{}}ND5k \\ LUMO Energy\end{tabular} & \begin{tabular}[c]{@{}l@{}}ND5k-lt \\ HOMO Energy\end{tabular} & \begin{tabular}[c]{@{}l@{}}ND5k-lt \\ LUMO Energy\end{tabular} \\ 
\hline
SOAP KRR     & 0.28 (0.41) & 0.35 (0.50) &  1.2 (1.7) & 0.66 (0.82) \\
enn-s2s      & 0.22 (0.35) & 0.23 (0.41) & 0.48 (0.85) & 0.26 (0.34) \\
SOAP-enn-s2s  & 0.22 (0.35) & 0.21 (0.39) & 2.2 (2.7) & 1.5 (1.7) \\
SOAP-PCA-enn-s2s & 0.18 (0.27) & \textbf{0.20 (0.37)} & 0.38 (0.63) & 0.25 (0.33) \\
SchNet     & 0.23 (0.35) & 0.22 (0.39) & 0.45 (0.70) & 0.26 (0.34) \\
PaiNN avg pool & \textbf{0.16 (0.27)} & \textbf{0.19 (0.32)} & \textbf{0.34 (0.52)} & \textbf{0.23 (0.31)} \\
PaiNN sum pool & 0.18 (0.29) & \textbf{0.20 (0.35)} & 2.6 (3.1) & 0.53 (0.71)
\end{tabular}
\label{tab:ML}
\end{table}

\begin{figure}[H]
 \raggedright
 \includegraphics[width = 0.95\textwidth]{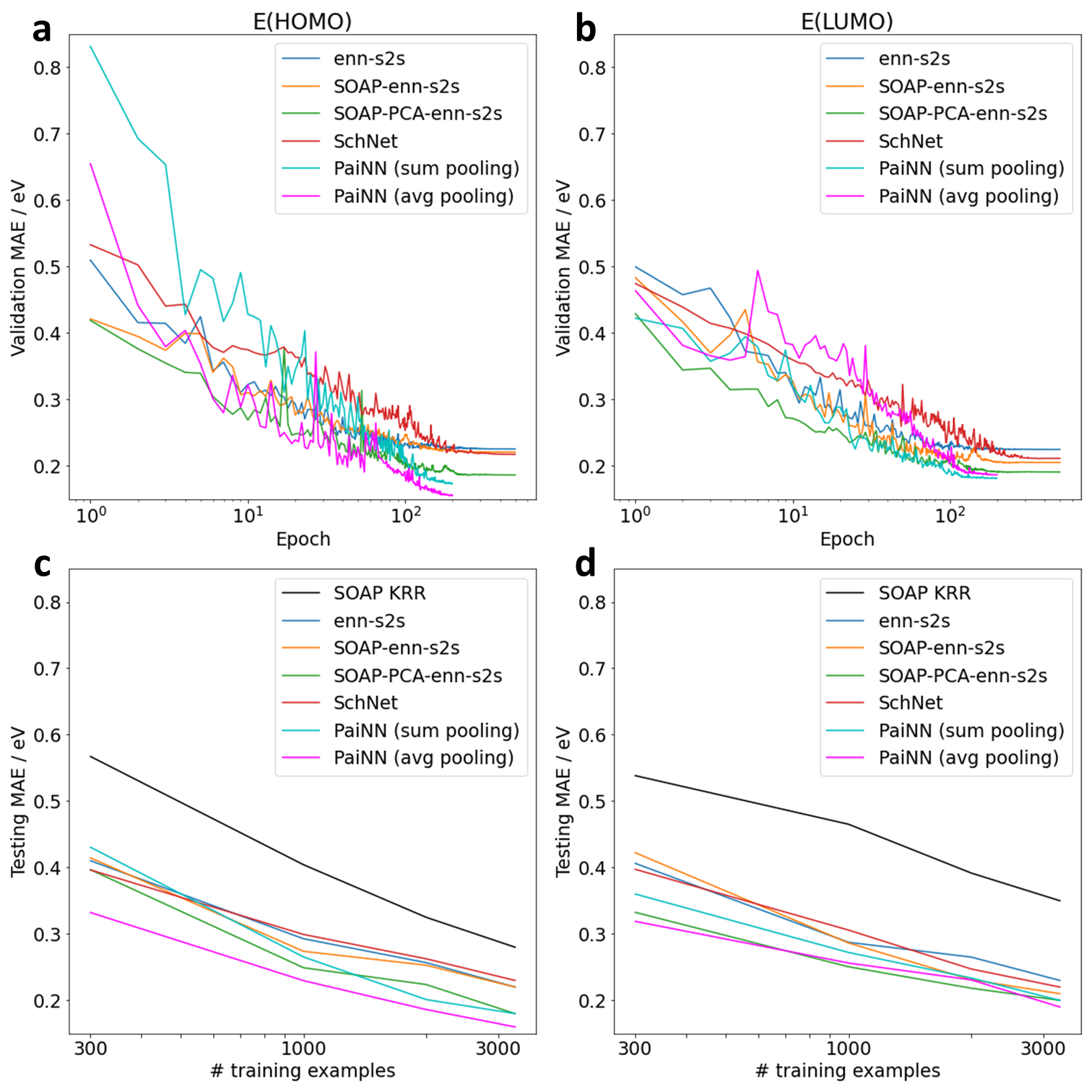}
    \caption{Top: Validation learning curves of the graph neural networks on ND5k (a) HOMO and (b) LUMO energy training, averaged over all training runs (cf. table \ref{tab:ML}). Mean absolute validation error (MAE) in eV is plotted against the number of epochs (log scale). Bottom: Learning curves with respect to the training set size on ND5k (c) HOMO and (d) LUMO energy training. Mean absolute testing error (MAE) in eV is plotted against the number of training examples (log scale).}
    \label{fig:LCs}
\end{figure} 

Of all ML methods employed for ND5k learning, the SOAP KRR yields the highest errors (MAEs aound 0.3 eV), and the GNNs enn-s2s and SchNet achieve slightly better results (MAEs around 0.22 eV). The LUMO energy predictions of the SOAP-enn-s2s variant are improved by 0.02 eV MAE compared to the plain enn-s2s, and the SOAP-PCA-enn-s2s yields a further small improvement of 0.04 eV and 0.01 eV for HOMO and LUMO energies, respectively. The best results are obtained from the PaiNN model that uses average pooling, with MAEs of 0.16 eV and 0.19 eV for HOMO and LUMO energy predictions, respectively. The second best results for the interpolation task achieved by the PaiNN model with sum pooling and the SOAP-PCA-enn-s2s.

The incorporation of additional structural information into the enn-s2s graph nodes via plain SOAP descriptors does not significantly affect the learning performance on the interpolation task. For extrapolation to ND5k-lt, the MAEs are even strongly increased to $>$ 1 eV. The removal of redundant information from the SOAP vectors via PCA then notably improves the results by increasing learning efficiency while making the model less prone to overfitting: In the SOAP-PCA-enn-s2s dimensionality reduction, 99.99 \% of the SOAP vectors' variance was retained after reducing their size by over 90 \%, thereby also improving the computational efficiency of the network. The SOAP vectors carry information on all possible three-body-interactions between all elements present in the dataset, many of which are obsolete for most structures, yielding highly sparse descriptors. Thus, in accordance with earlier studies,\cite{parsaeifard_assessment_2021, darby_compressing_2022} the unedited SOAP vectors contain a large amount of redundant information, which probably also limits the effectiveness of the SOAP KRR model. However, finally, the PaiNN architecture is best able to efficiently exploit the structural information within the ND5k dataset, and expressive enough to model the highly non-linear structure-property relationships, yielding the lowest MAEs. 

As can be seen in figure \ref{fig:LCs}a and b, the learning curves of all graph neural networks are very similar to each other. For PaiNN, the use of average pooling, which directly encodes size-intensivity into the architecture, accelerates the learning over sum pooling. Interestingly, PaiNN initially scores higher errors than the other networks, but then converges to the best final result.
The learning curves with respect to the number of training examples are depicted in figure \ref{fig:LCs}c and d. Kernel regression tends to be superior when less training data is available, however, the actual performance always depends on the specific learning task.\cite{faber_prediction_2017, yaghoobi_machine_2022} Here, the SOAP KRR is constantly outperformed by all GNNs, even for small training set sizes. Apparently, the GNN architectures are generally superior to the SOAP KRR approach for the learning task at hand.
For HOMO energy predictions, PaiNN with average pooling constantly yields the best results, however, for LUMO energy predictions it gets slightly outperformed by the SOAP-PCA-enn-s2s for intermediate training set sizes (1000 and 2000 datapoints). Interestingly, the SOAP-enn-s2s initially gets outperformed by the plain enn-s2s, indicating that the large SOAP-enhanced architecture needs more data to learn effectively.
Generally, the enn-s2s curves clearly substantiate that the incorporation of SOAP features into the architecture and the PCA-reduction of the SOAP descriptors both accelerate learning and improve the final result. 

We also compared the performance of the SOAP-PCA-enn-s2s model to the plain enn-s2s on two different datasets, QM9 (134k small organic molecules)\cite{ruddigkeit_enumeration_2012, ramakrishnan_quantum_2014} and OE62 (62k diverse medium size organic molecules)\cite{stuke_atomic_2020} for learning frontier orbital energies. For both datasets, the enn-s2s architecture was again fine-tuned using a small grid-based hyperparameter search as described in the supporting information. On the QM9 dataset, the plain enn-s2s scored MAEs of 84 meV and 86 meV for HOMO and LUMO energies, respectively. The error was reduced to 60 meV and 79 meV, respectively, when using the SOAP-PCA variant. We note here that, while this is a substantial improvement, the errors are still higher than state of the are accuracy (< 25 meV MAEs)\cite{klicpera2020fast, liu_transferable_2021} and can probably be further reduced by a more elaborate tuning of hyperparameters. On the OE62 dataset, MAEs of 145 meV and 141 meV were obtained for HOMO and LUMO energies, respectively, using the plain enn-s2s. When using the SOAP-PCA variant, the MAEs decreased to 115 meV for both targets. Similar MAEs for this task were reported by Ramahan et al. when using a variant of the enn-s2s architecture (E$_{HOMO}$: 148 meV, E$_{LUMO}$: 145 meV).\cite{rahaman_deep_2020} In this work, the authors also used different variations of hybrid learning architectures to combine the enn-s2s GNN with molecular descriptors, however, the lowest MAEs from these architectures (E$_{HOMO}$: 131 meV, E$_{LUMO}$: 130 meV) are higher than those obtained from the SOAP-PCA-enn-s2s presented here.\cite{rahaman_deep_2020}

The errors reported here for the ND5k dataset are much higher than those reported for other quantum chemistry datasets: Common frontier orbital energy MAEs are $<$ 0.05 eV on QM9\cite{schutt_equivariant_2021} and $<$ 0.15 eV on OE62.\cite{rahaman_deep_2020} The higher errors arise from the high complexity and comparatively low size of the dataset, but are also a result of the conservative split (66 \% training, 17 \% validation, 17 \% test) that was applied. 

For the case of extrapolating predictions to larger nanodiamonds to the ND5k-lt set after training on the ND5k dataset, the SOAP KRR approach again performs rather poorly. Also, the SOAP-enn-s2s network scores high errors, which is probably due to overfitting after the incorporation of large sparse SOAP vectors into the learning architecture. However, also the prediction errors of PaiNN with sum pooling strongly deteriorate, especially for the prediction of HOMO energies (MAE of 2.6 eV). On the other hand, the less expressive SchNet and enn-s2s architectures as well as PaiNN with average pooling are much more robust in their predictions. Here, the average pooling PaiNN achieves the best results, with MAEs of 0.34 and 0.23 eV for HOMO and LUMO energy predictions, respectively. 

The PaiNN model that uses sum pooling scores the second best results on the interpolation task, together with SOAP-PCA-enn-s2s. This is surprising, because frontier orbital energies are intensive properties, and sum pooling is usually only advantageous for learning extensive properties. In this case, however, as a result of the quantum size effect, the HOMO energies and the total numbers of atoms n$_{\mathrm{atoms}}$ of the ND5k structures are strongly correlated: The Pearson correlation coefficient $r_c$ between E(HOMO) and n$_{\mathrm{atoms}}$ is $r_c$ = 0.47, while for E(LUMO) and n$_{\mathrm{atoms}}$ we find $r_c$ = 0.15.\cite{chang_quantum_1999, willey_molecular_2005, bolker_quantum_2013, pearson1903laws}. This energy dependence saturates for large n$_{\mathrm{atoms}}$, as illustrated in figures S4 and S5 in the SI.

The saturation of the quantum confinement effect leads to higher prediction errors for the structures outside the ND5k training range when the ML models intrinsically assume size extensivity by means of sum pooling. This effect is more pronounced for HOMO energy predictions, where the size correlation within ND5k is very strong and breaks down for the larger structures (see figure S4). The same effect, despite less pronounced, can be identified for the correlation of n$_atoms$ and LUMO energies. Especially the PaiNN with sum pooling gets fooled by the correlation within the ND5k dataset and its prediction errors increase drastically, correlating with increasing nanodiamond sizes.


\end{justify}

\clearpage

\section{Conclusions}

\begin{justify}

In this study, we introduced the ND5k dataset, containing DFTB-optimized structures and DFT/PBE0-computed frontier orbital energies of 5,089 diamondoids and nanodiamonds. The dataset is based on 17 H-terminated, undoped base structures, covered by either one of five surface species (H, F, OH, NH$_2$, F/H, OH/H, NH$_2$/H) and (co-) doped with up to two dopant atoms (B, N, Si, P). From this dataset, we extracted a subset of 45 structures with frontier orbital alignments that make them suitable for use as photocatalysts in solar light-driven reduction reactions. The structures from this subset suggest that P-doped, H- and NH$_2$-covered nanodiamonds are promising candidate materials for this field. For further applications, such as electrode materials and sensors, the ND5k dataset provides both a source of pre-computed candidate structures and a basis for ML-guided design.

Furthermore, we tested six ML models for predicting the frontier orbital energies of the ND5k structures, and we evaluated their performances on extrapolating their predictions to larger nanodiamonds. For this, we exchanged the generic atom-wise node embeddings of an enn-s2s by their plain (SOAP-enn-s2s) or PCA-reduced (SOAP-PCA-enn-s2s) SOAP atomic descriptors. We found the latter to clearly outperform the other two enn-s2s variants, not only on ND5k, but also on the QM9 and OE62 datasets. The PaiNN GNN outperformed all other MD models on both the classical prediction task on ND5k, and on the extrapolation task on ND5k-lt. The second best overall results were obtained from the SOAP-PCA-enn-s2s.

The general approach used in the SOAP-PCA-enn-s2s architecture allows for the straight-forward incorporation of structural features into any GNN. Future investigations towards more efficient and robust learning algorithms making use of this approach may target different combinations of atomic descriptors, dimensionality reduction techniques, and GNNs.

\end{justify}

\section{Acknowledgements}

\begin{justify}

We thank Prof. Gabriel Bester for providing pre-optimized structures of nanodiamonds, and Prof. Philipp Marquetand, Dr. Félix Musil and Dr. Kristof T. Schütt for helpful discussions.
TK, JD and FN acknowledge support from the Helmholtz Einstein International Berlin Research School in Data Science (HEIBRiDS). FN acknowledges support from European Commission (ERC CoG 772230), The Berlin Mathematics center MATH+ (AA2-8) and the Berlin Institute for the Foundations of Learning and Data (BIFOLD).
Computing resources were kindly provided by the Freie Universität Berlin hpc cluster Curta\cite{bennett_curta_2020} and by the Helmholtz-Zentrum Dresden-Rossendorf.

\end{justify}

\section{Data Availability Statement}
\label{sec:data_availability}

\begin{justify}

The ND5k dataset, the test set of larger nanodiamonds, and the implementation of the PCA-SOAP-enn-s2s can be found online at https://github.com/ThorrenKirschbaum/ND5k 
The ND5k dataset will be uploaded to the NOMAD repository and the link inserted here in the final stage of publication.

\end{justify}

\bibliography{references}

\begin{thebibliography}{100}

\bibitem{roduner_size_2006}
Emil Roduner.
\newblock Size matters: why nanomaterials are different.
\newblock {\em Chem. Soc. Rev.}, 35(7):583--592, 2006.
\newblock Publisher: Royal Society of Chemistry.

\bibitem{baig_nanomaterials_2021}
Nadeem Baig, Irshad Kammakakam, and Wail Falath.
\newblock Nanomaterials: a review of synthesis methods, properties, recent
  progress, and challenges.
\newblock {\em Mater. Adv.}, 2(6):1821--1871, 2021.
\newblock Publisher: Royal Society of Chemistry.

\bibitem{mochalin_properties_2012}
Vadym~N. Mochalin, Olga Shenderova, Dean Ho, and Yury Gogotsi.
\newblock The properties and applications of nanodiamonds.
\newblock {\em Nat. Nanotechnol.}, 7(1):11--23, January 2012.

\bibitem{nunn_nanodiamond_2017}
Nicholas Nunn, Marco Torelli, Gary McGuire, and Olga Shenderova.
\newblock Nanodiamond: {A} high impact nanomaterial.
\newblock {\em Curr. Opin. Solid State Mater. Sci.}, 21(1):1--9, February 2017.

\bibitem{zhu_photo-illuminated_2013}
Di~Zhu, Linghong Zhang, Rose~E. Ruther, and Robert~J. Hamers.
\newblock Photo-illuminated diamond as a solid-state source of solvated
  electrons in water for nitrogen reduction.
\newblock {\em Nat. Mater.}, 12(9):836--841, September 2013.
\newblock Number: 9 Publisher: Nature Publishing Group.

\bibitem{zhang_selective_2014}
Linghong Zhang, Di~Zhu, Gilbert~M. Nathanson, and Robert~J. Hamers.
\newblock Selective {Photoelectrochemical} {Reduction} of {Aqueous} {CO2} to
  {CO} by {Solvated} {Electrons}.
\newblock {\em Angew. Chem. Int. Ed.}, 53(37):9746--9750, 2014.
\newblock \_eprint:
  https://onlinelibrary.wiley.com/doi/pdf/10.1002/anie.201404328.

\bibitem{hamers_photoemission_2014}
R.~J. Hamers, J.~A. Bandy, D.~Zhu, and L.~Zhang.
\newblock Photoemission from diamond films and substrates into water: dynamics
  of solvated electrons and implications for diamond photoelectrochemistry.
\newblock {\em Faraday Discuss.}, 172(0):397--411, November 2014.

\bibitem{zhang_photocatalytic_2017}
Linghong Zhang and Robert~J. Hamers.
\newblock Photocatalytic reduction of {CO2} to {CO} by diamond nanoparticles.
\newblock {\em Diam. Relat. Mater.}, 78:24--30, September 2017.

\bibitem{lai_surface_2012}
Lin Lai and Amanda~S. Barnard.
\newblock Surface phase diagram and thermodynamic stability of
  functionalisation of nanodiamonds.
\newblock {\em J. Mater. Chem.}, 22(33):16774--16780, July 2012.
\newblock Publisher: The Royal Society of Chemistry.

\bibitem{petit_valence_2015}
Tristan Petit, Mika Pflüger, Daniel Tolksdorf, Jie Xiao, and Emad~F. Aziz.
\newblock Valence holes observed in nanodiamonds dispersed in water.
\newblock {\em Nanoscale}, 7(7):2987--2991, 2015.

\bibitem{petit_unusual_2017}
Tristan Petit, Ljiljana Puskar, Tatiana Dolenko, Sneha Choudhury, Eglof Ritter,
  Sergey Burikov, Kirill Laptinskiy, Quentin Brzustowski, Ulrich Schade, Hayato
  Yuzawa, Masanari Nagasaka, Nobuhiro Kosugi, Magdalena Kurzyp, Amélie
  Venerosy, Hugues Girard, Jean-Charles Arnault, Eiji Osawa, Nicholas Nunn,
  Olga Shenderova, and Emad~F. Aziz.
\newblock Unusual {Water} {Hydrogen} {Bond} {Network} around {Hydrogenated}
  {Nanodiamonds}.
\newblock {\em J. Phys. Chem. C}, 121(9):5185--5194, March 2017.

\bibitem{choudhury_combining_2018}
Sneha Choudhury, Benjamin Kiendl, Jian Ren, Fang Gao, Peter Knittel, Christoph
  Nebel, Amélie Venerosy, Hugues Girard, Jean-Charles Arnault, Anke Krueger,
  Karin Larsson, and Tristan Petit.
\newblock Combining nanostructuration with boron doping to alter sub band gap
  acceptor states in diamond materials.
\newblock {\em J. Mater. Chem. A}, 6(34):16645--16654, August 2018.
\newblock Publisher: The Royal Society of Chemistry.

\bibitem{feigl_classifying_2019}
C.~A. Feigl, B.~Motevalli, A.~J. Parker, B.~Sun, and A.~S. Barnard.
\newblock Classifying and predicting the electron affinity of diamond
  nanoparticles using machine learning.
\newblock {\em Nanoscale Horiz.}, 4(4):983--990, June 2019.
\newblock Publisher: The Royal Society of Chemistry.

\bibitem{kirschbaum_effects_2022}
Thorren Kirschbaum, Tristan Petit, Joachim Dzubiella, and Annika Bande.
\newblock Effects of oxidative adsorbates and cluster formation on the
  electronic structure of nanodiamonds.
\newblock {\em J. Comput. Chem.}, 43(13):923--929, 2022.
\newblock \_eprint: https://onlinelibrary.wiley.com/doi/pdf/10.1002/jcc.26849.

\bibitem{teunissen_tuning_2017}
Jos~L. Teunissen, Frank De~Proft, and Freija De~Vleeschouwer.
\newblock Tuning the {HOMO}–{LUMO} {Energy} {Gap} of {Small} {Diamondoids}
  {Using} {Inverse} {Molecular} {Design}.
\newblock {\em J. Chem. Theory Comput.}, 13(3):1351--1365, March 2017.

\bibitem{zhang_hybrid_2018}
Ting Zhang, Gang-Qin Liu, Weng-Hang Leong, Chu-Feng Liu, Man-Hin Kwok, To~Ngai,
  Ren-Bao Liu, and Quan Li.
\newblock Hybrid nanodiamond quantum sensors enabled by volume phase
  transitions of hydrogels.
\newblock {\em Nat. Commun.}, 9(1):3188, August 2018.
\newblock Number: 1 Publisher: Nature Publishing Group.

\bibitem{wang_nanodiamonds_2019}
Hongxia Wang and Yi~Cui.
\newblock Nanodiamonds for energy.
\newblock {\em Carbon energy}, 1(1):13--18, 2019.
\newblock \_eprint: https://onlinelibrary.wiley.com/doi/pdf/10.1002/cey2.9.

\bibitem{liu_nanodiamond-enabled_2020}
Yen-Yiu Liu, Be-Ming Chang, and Huan-Cheng Chang.
\newblock Nanodiamond-enabled biomedical imaging.
\newblock {\em Nanomedicine}, 15(16):1599--1616, July 2020.
\newblock Publisher: Future Medicine.

\bibitem{chang_quantum_1999}
Y.~K. Chang, H.~H. Hsieh, W.~F. Pong, M.-H. Tsai, F.~Z. Chien, P.~K. Tseng,
  L.~C. Chen, T.~Y. Wang, K.~H. Chen, D.~M. Bhusari, J.~R. Yang, and S.~T. Lin.
\newblock Quantum {Confinement} {Effect} in {Diamond} {Nanocrystals} {Studied}
  by {X}-{Ray}-{Absorption} {Spectroscopy}.
\newblock {\em Phys. Rev. Lett.}, 82(26):5377--5380, June 1999.

\bibitem{willey_molecular_2005}
T.~M. Willey, C.~Bostedt, T.~van Buuren, J.~E. Dahl, S.~G. Liu, R.~M.~K.
  Carlson, L.~J. Terminello, and T.~Möller.
\newblock Molecular {Limits} to the {Quantum} {Confinement} {Model} in
  {Diamond} {Clusters}.
\newblock {\em Phys. Rev. Lett.}, 95(11):113401, September 2005.
\newblock Publisher: American Physical Society.

\bibitem{stehlik_size_2015}
Stepan Stehlik, Marian Varga, Martin Ledinsky, Vit Jirasek, Anna Artemenko,
  Halyna Kozak, Lukas Ondic, Viera Skakalova, Giacomo Argentero, Timothy
  Pennycook, Jannik~C. Meyer, Antonin Fejfar, Alexander Kromka, and Bohuslav
  Rezek.
\newblock Size and {Purity} {Control} of {HPHT} {Nanodiamonds} down to 1 nm.
\newblock {\em J. Phys. Chem. C}, 119(49):27708--27720, December 2015.
\newblock Publisher: American Chemical Society.

\bibitem{ferro_physicochemical_2003}
Sergio Ferro and Achille De~Battisti.
\newblock Physicochemical {Properties} of {Fluorinated} {Diamond} {Electrodes}.
\newblock {\em J. Phys. Chem. B}, 107(31):7567--7573, August 2003.
\newblock Publisher: American Chemical Society.

\bibitem{wang_comparison_2009}
Mei Wang, Nathalie Simon, Claudia Decorse-Pascanut, Muriel Bouttemy, Arnaud
  Etcheberry, Musen Li, Rabah Boukherroub, and Sabine Szunerits.
\newblock Comparison of the chemical composition of boron-doped diamond
  surfaces upon different oxidation processes.
\newblock {\em Electrochim. Acta}, 54(24):5818--5824, October 2009.

\bibitem{brown_controlling_2014}
Noam Brown and Oded Hod.
\newblock Controlling the {Electronic} {Properties} of {Nanodiamonds} via
  {Surface} {Chemical} {Functionalization}: {A} {DFT} {Study}.
\newblock {\em J. Phys. Chem. C}, 118(10):5530--5537, March 2014.
\newblock Publisher: American Chemical Society.

\bibitem{larsson_effect_2018}
K.~Larsson and Y.~Tian.
\newblock Effect of surface termination on the reactivity of nano-sized diamond
  particle surfaces for bio applications.
\newblock {\em Carbon}, 134:244--254, August 2018.

\bibitem{pinault_n-type_2007}
M.~A. Pinault, J.~Barjon, T.~Kociniewski, F.~Jomard, and J.~Chevallier.
\newblock The n-type doping of diamond: {Present} status and pending questions.
\newblock {\em Physica B: Condens. Matter}, 401-402:51--56, December 2007.

\bibitem{williams_growth_2008}
O.~A. Williams, M.~Nesladek, M.~Daenen, S.~Michaelson, A.~Hoffman, E.~Osawa,
  K.~Haenen, and R.~B. Jackman.
\newblock Growth, electronic properties and applications of nanodiamond.
\newblock {\em Diam. Relat. Mater.}, 17(7):1080--1088, July 2008.

\bibitem{knittel_nanostructured_2020}
Peter Knittel, Franziska Buchner, Emina Hadzifejzovic, Christian Giese,
  Patricia Quellmalz, Robert Seidel, Tristan Petit, Boyan Iliev, Thomas J.~S.
  Schubert, Christoph~E. Nebel, and John~S. Foord.
\newblock Nanostructured {Boron} {Doped} {Diamond} {Electrodes} with
  {Increased} {Reactivity} for {Solar}-{Driven} {CO2} {Reduction} in {Room}
  {Temperature} {Ionic} {Liquids}.
\newblock {\em ChemCatChem}, 12(21):5548--5557, 2020.
\newblock \_eprint:
  https://onlinelibrary.wiley.com/doi/pdf/10.1002/cctc.202000938.

\bibitem{pichot_efficient_2008}
V.~Pichot, M.~Comet, E.~Fousson, C.~Baras, A.~Senger, F.~Le~Normand, and
  D.~Spitzer.
\newblock An efficient purification method for detonation nanodiamonds.
\newblock {\em Diam. Relat. Mater.}, 17(1):13--22, January 2008.

\bibitem{mikheev_low-power_2020}
Konstantin~G. Mikheev, Tatyana~N. Mogileva, Arseniy~E. Fateev, Nicholas~A.
  Nunn, Olga~A. Shenderova, and Gennady~M. Mikheev.
\newblock Low-{Power} {Laser} {Graphitization} of {High} {Pressure}—{High}
  {Temperature} {Nanodiamond} {Films}.
\newblock {\em Appl. Sci.}, 10(9):3329, January 2020.
\newblock Number: 9 Publisher: Multidisciplinary Digital Publishing Institute.

\bibitem{behler_nanodiamond-polymer_2009}
Kristopher~D. Behler, Antonella Stravato, Vadym Mochalin, Guzeliya Korneva,
  Gleb Yushin, and Yury Gogotsi.
\newblock Nanodiamond-{Polymer} {Composite} {Fibers} and {Coatings}.
\newblock {\em ACS Nano}, 3(2):363--369, February 2009.
\newblock Publisher: American Chemical Society.

\bibitem{bandy_photocatalytic_2016}
Jason~A. Bandy, Di~Zhu, and Robert~J. Hamers.
\newblock Photocatalytic reduction of nitrogen to ammonia on diamond thin films
  grown on metallic substrates.
\newblock {\em Diam. Relat. Mater.}, 64:34--41, April 2016.

\bibitem{etemadi_performance_2017}
Habib Etemadi, Reza Yegani, and Valiollah Babaeipour.
\newblock Performance evaluation and antifouling analyses of cellulose
  acetate/nanodiamond nanocomposite membranes in water treatment: {ARTICLE}.
\newblock {\em J. Appl. Polym. Sci.}, 134(21), June 2017.

\bibitem{petit-dominguez_synergistic_2018}
María~Dolores Petit-Domínguez, Carmen Quintana, Luis Vázquez, María del
  Pozo, Isabel Cuadrado, Ana~María Parra-Alfambra, and Elena Casero.
\newblock Synergistic effect of {MoS2} and diamond nanoparticles in
  electrochemical sensors: determination of the anticonvulsant drug valproic
  acid.
\newblock {\em Microchim. Acta}, 185(7):334, June 2018.

\bibitem{dral_quantum_2020}
Pavlo~O. Dral.
\newblock Quantum {Chemistry} in the {Age} of {Machine} {Learning}.
\newblock {\em J. Phys. Chem. Lett.}, 11(6):2336--2347, March 2020.
\newblock Publisher: American Chemical Society.

\bibitem{foulkes1989tight}
W~Matthew~C Foulkes and Roger Haydock.
\newblock Tight-binding models and density-functional theory.
\newblock {\em Phys. Rev. B}, 39(17):12520, 1989.

\bibitem{li_machine-learning_2018}
Zheng Li, Noushin Omidvar, Wei~Shan Chin, Esther Robb, Amanda Morris, Luke
  Achenie, and Hongliang Xin.
\newblock Machine-{Learning} {Energy} {Gaps} of {Porphyrins} with {Molecular}
  {Graph} {Representations}.
\newblock {\em J. Phys. Chem. A}, 122(18):4571--4578, May 2018.
\newblock Publisher: American Chemical Society.

\bibitem{lee_insights_2019}
Min-Hsuan Lee.
\newblock Insights from {Machine} {Learning} {Techniques} for {Predicting} the
  {Efficiency} of {Fullerene} {Derivatives}-{Based} {Ternary} {Organic} {Solar}
  {Cells} at {Ternary} {Blend} {Design}.
\newblock {\em Adv. Energy Mater.}, 9(26):1900891, 2019.
\newblock \_eprint:
  https://onlinelibrary.wiley.com/doi/pdf/10.1002/aenm.201900891.

\bibitem{padula_combining_2019}
Daniele Padula, Jack D. Simpson, and Alessandro Troisi.
\newblock Combining electronic and structural features in machine learning
  models to predict organic solar cells properties.
\newblock {\em Mater. Horizons}, 6(2):343--349, 2019.
\newblock Publisher: Royal Society of Chemistry.

\bibitem{lee_machine_2020}
Min-Hsuan Lee.
\newblock A {Machine} {Learning}–{Based} {Design} {Rule} for {Improved}
  {Open}-{Circuit} {Voltage} in {Ternary} {Organic} {Solar} {Cells}.
\newblock {\em Adv. Intell. Syst.}, 2(1):1900108, 2020.
\newblock \_eprint:
  https://onlinelibrary.wiley.com/doi/pdf/10.1002/aisy.201900108.

\bibitem{meftahi_machine_2020}
Nastaran Meftahi, Mykhailo Klymenko, Andrew~J. Christofferson, Udo Bach,
  David~A. Winkler, and Salvy~P. Russo.
\newblock Machine learning property prediction for organic photovoltaic
  devices.
\newblock {\em npj Comput. Mater.}, 6(1):1--8, November 2020.
\newblock Number: 1 Publisher: Nature Publishing Group.

\bibitem{lee_identifying_2022}
Min-Hsuan Lee.
\newblock Identifying correlation between the open-circuit voltage and the
  frontier orbital energies of non-fullerene organic solar cells based on
  interpretable machine-learning approaches.
\newblock {\em Sol. Energy}, 234:360--367, March 2022.

\bibitem{zhang_high-efficiency_2022}
Qi~Zhang, Yu~Jie Zheng, Wenbo Sun, Zeping Ou, Omololu Odunmbaku, Meng Li,
  Shanshan Chen, Yongli Zhou, Jing Li, Bo~Qin, and Kuan Sun.
\newblock High-{Efficiency} {Non}-{Fullerene} {Acceptors} {Developed} by
  {Machine} {Learning} and {Quantum} {Chemistry}.
\newblock {\em Adv. Sci.}, 9(6):2104742, 2022.
\newblock \_eprint:
  https://onlinelibrary.wiley.com/doi/pdf/10.1002/advs.202104742.

\bibitem{storm_machine_2022}
Freja~E. Storm, Linnea~M. Folkmann, Thorsten Hansen, and Kurt~V. Mikkelsen.
\newblock Machine learning the frontier orbital energies of {SubPc} based
  triads.
\newblock {\em J. Mol. Model.}, 28(10):313, September 2022.

\bibitem{pereira_machine_2017}
Florbela Pereira, Kaixia Xiao, Diogo A. R.~S. Latino, Chengcheng Wu, Qingyou
  Zhang, and Joao Aires-de Sousa.
\newblock Machine {Learning} {Methods} to {Predict} {Density} {Functional}
  {Theory} {B3LYP} {Energies} of {HOMO} and {LUMO} {Orbitals}.
\newblock {\em J. Chem. Inf. Model.}, 57(1):11--21, January 2017.
\newblock Publisher: American Chemical Society.

\bibitem{mchang_hammett_2019}
Alexander M.~Chang, Jessica G.~Freeze, and Victor S.~Batista.
\newblock Hammett neural networks: prediction of frontier orbital energies of
  tungsten–benzylidyne photoredox complexes.
\newblock {\em Chem. Sci.}, 10(28):6844--6854, 2019.
\newblock Publisher: Royal Society of Chemistry.

\bibitem{olsthoorn_band_2019}
Bart Olsthoorn, R.~Matthias Geilhufe, Stanislav~S. Borysov, and Alexander~V.
  Balatsky.
\newblock Band {Gap} {Prediction} for {Large} {Organic} {Crystal} {Structures}
  with {Machine} {Learning}.
\newblock {\em Adv. Quantum Technol.}, 2(7-8):1900023, 2019.
\newblock \_eprint:
  https://onlinelibrary.wiley.com/doi/pdf/10.1002/qute.201900023.

\bibitem{rahaman_deep_2020}
Obaidur Rahaman and Alessio Gagliardi.
\newblock Deep {Learning} {Total} {Energies} and {Orbital} {Energies} of
  {Large} {Organic} {Molecules} {Using} {Hybridization} of {Molecular}
  {Fingerprints}.
\newblock {\em J. Chem. Inf. Model.}, 60(12):5971--5983, December 2020.
\newblock Publisher: American Chemical Society.

\bibitem{woon_relating_2021}
Kai~Lin Woon, Zhao~Xian Chong, Azhar Ariffin, and Chee~Seng Chan.
\newblock Relating molecular descriptors to frontier orbital energy levels,
  singlet and triplet excited states of fused tricyclics using machine
  learning.
\newblock {\em J. Mol. Graph. Model.}, 105:107891, June 2021.

\bibitem{ye_assessment_2022}
Zong-Rong Ye, Sheng-Hsuan Hung, Berlin Chen, and Ming-Kang Tsai.
\newblock Assessment of {Predicting} {Frontier} {Orbital} {Energies} for
  {Small} {Organic} {Molecules} {Using} {Knowledge}-{Based} and {Structural}
  {Information}.
\newblock {\em ACS Eng. Au}, 2(4):360--368, August 2022.
\newblock Publisher: American Chemical Society.

\bibitem{duan_machine_2021}
Chenru Duan, Shuxin Chen, Michael G. Taylor, Fang Liu, and Heather J.~Kulik.
\newblock Machine learning to tame divergent density functional approximations:
  a new path to consensus materials design principles.
\newblock {\em Chem. Sci.}, 12(39):13021--13036, 2021.
\newblock Publisher: Royal Society of Chemistry.

\bibitem{mazouin_selected_2022}
Bernard Mazouin, Alexandre Alain Schöpfer, and O.~Anatole~von Lilienfeld.
\newblock Selected machine learning of {HOMO}–{LUMO} gaps with improved
  data-efficiency.
\newblock {\em Mater. Adv.}, 2022.
\newblock Publisher: Royal Society of Chemistry.

\bibitem{nigam_unified_2022}
Jigyasa Nigam, Guillaume Fraux, and Michele Ceriotti.
\newblock Unified theory of atom-centered representations and graph
  convolutional machine-learning schemes.
\newblock {\em arXiv:2202.01566 [physics, stat]}, February 2022.
\newblock arXiv: 2202.01566.

\bibitem{behler_generalized_2007}
Jorg Behler and Michele Parrinello.
\newblock Generalized {Neural}-{Network} {Representation} of
  {High}-{Dimensional} {Potential}-{Energy} {Surfaces}.
\newblock {\em Phys. Rev. Lett.}, page~4, 2007.

\bibitem{bartok_representing_2013}
Albert~P. Bartók, Risi Kondor, and Gábor Csányi.
\newblock On representing chemical environments.
\newblock {\em Phys. Rev. B}, 87(18):184115, May 2013.

\bibitem{parsaeifard_assessment_2021}
Behnam Parsaeifard, Deb~Sankar De, Anders~S. Christensen, Felix~A. Faber, Emir
  Kocer, Sandip De, Jörg Behler, O.~Anatole von Lilienfeld, and Stefan
  Goedecker.
\newblock An assessment of the structural resolution of various fingerprints
  commonly used in machine learning.
\newblock {\em Mach. Learn.: Sci. Technol.}, 2(1):015018, March 2021.
\newblock Publisher: IOP Publishing.

\bibitem{musil_physics-inspired_2021}
Felix Musil, Andrea Grisafi, Albert~P. Bartók, Christoph Ortner, Gábor
  Csányi, and Michele Ceriotti.
\newblock Physics-{Inspired} {Structural} {Representations} for {Molecules} and
  {Materials}.
\newblock {\em Chem. Rev.}, 121(16):9759--9815, August 2021.
\newblock Publisher: American Chemical Society.

\bibitem{kipf2016}
Thomas~N Kipf and Max Welling.
\newblock Semi-supervised classification with graph convolutional networks.
\newblock {\em arXiv preprint arXiv:1609.02907}, 2016.

\bibitem{gilmer_neural_2017}
Justin Gilmer, Samuel~S. Schoenholz, Patrick~F. Riley, Oriol Vinyals, and
  George~E. Dahl.
\newblock Neural {Message} {Passing} for {Quantum} {Chemistry}.
\newblock {\em arXiv:1704.01212 [cs]}, June 2017.
\newblock arXiv: 1704.01212.

\bibitem{schutt_schnet_2017}
Kristof Schütt, Pieter-Jan Kindermans, Huziel~Enoc Sauceda~Felix, Stefan
  Chmiela, Alexandre Tkatchenko, and Klaus-Robert Müller.
\newblock {SchNet}: {A} continuous-filter convolutional neural network for
  modeling quantum interactions.
\newblock In {\em Advances in {Neural} {Information} {Processing} {Systems}},
  volume~30. Curran Associates, Inc., 2017.

\bibitem{schutt_quantum-chemical_2017}
Kristof~T. Schütt, Farhad Arbabzadah, Stefan Chmiela, Klaus~R. Müller, and
  Alexandre Tkatchenko.
\newblock Quantum-chemical insights from deep tensor neural networks.
\newblock {\em Nat. Commun.}, 8(1):13890, January 2017.
\newblock Number: 1 Publisher: Nature Publishing Group.

\bibitem{thomas2018}
Nathaniel Thomas, Tess Smidt, Steven Kearnes, Lusann Yang, Li~Li, Kai Kohlhoff,
  and Patrick Riley.
\newblock Tensor field networks: Rotation- and translation-equivariant neural
  networks for 3d point clouds.
\newblock {\em arXiv preprint arXiv:1802.08219}, 2018.

\bibitem{schutt_equivariant_2021}
Kristof Schütt, Oliver Unke, and Michael Gastegger.
\newblock Equivariant message passing for the prediction of tensorial
  properties and molecular spectra.
\newblock In {\em Proceedings of the 38th {International} {Conference} on
  {Machine} {Learning}}, pages 9377--9388. PMLR, July 2021.
\newblock ISSN: 2640-3498.

\bibitem{miller_relevance_2020}
Benjamin~Kurt Miller, Mario Geiger, Tess~E. Smidt, and Frank Noé.
\newblock Relevance of {Rotationally} {Equivariant} {Convolutions} for
  {Predicting} {Molecular} {Properties}.
\newblock Technical Report arXiv:2008.08461, arXiv, November 2020.
\newblock arXiv:2008.08461 [physics, stat].

\bibitem{batzner_e3-equivariant_2022}
Simon Batzner, Albert Musaelian, Lixin Sun, Mario Geiger, Jonathan~P. Mailoa,
  Mordechai Kornbluth, Nicola Molinari, Tess~E. Smidt, and Boris Kozinsky.
\newblock E(3)-{Equivariant} {Graph} {Neural} {Networks} for {Data}-{Efficient}
  and {Accurate} {Interatomic} {Potentials}.
\newblock {\em Nat. Commun.}, 13(1):2453, December 2022.
\newblock arXiv:2101.03164 [cond-mat, physics:physics].

\bibitem{batatia2022}
Ilyes Batatia, D{\'a}vid~P{\'e}ter Kov{\'a}cs, Gregor~NC Simm, Christoph
  Ortner, and G{\'a}bor Cs{\'a}nyi.
\newblock Mace: Higher order equivariant message passing neural networks for
  fast and accurate force fields.
\newblock {\em arXiv preprint arXiv:2206.07697}, 2022.

\bibitem{geiger2022}
Mario Geiger and Tess Smidt.
\newblock e3nn: Euclidean neural networks.
\newblock {\em arXiv preprint arXiv:2207.09453}, 2022.

\bibitem{blum_970_2009}
Lorenz~C. Blum and Jean-Louis Reymond.
\newblock 970 {Million} {Druglike} {Small} {Molecules} for {Virtual}
  {Screening} in the {Chemical} {Universe} {Database} {GDB}-13.
\newblock {\em J. Am. Chem. Soc.}, 131(25):8732--8733, July 2009.
\newblock Publisher: American Chemical Society.

\bibitem{ruddigkeit_enumeration_2012}
Lars Ruddigkeit, Ruud van Deursen, Lorenz~C. Blum, and Jean-Louis Reymond.
\newblock Enumeration of 166 {Billion} {Organic} {Small} {Molecules} in the
  {Chemical} {Universe} {Database} {GDB}-17.
\newblock {\em J. Chem. Inf. Model.}, 52(11):2864--2875, November 2012.
\newblock Publisher: American Chemical Society.

\bibitem{montavon_machine_2013}
Grégoire Montavon, Matthias Rupp, Vivekanand Gobre, Alvaro Vazquez-Mayagoitia,
  Katja Hansen, Alexandre Tkatchenko, Klaus-Robert Müller, and O.~Anatole von
  Lilienfeld.
\newblock Machine learning of molecular electronic properties in chemical
  compound space.
\newblock {\em New J. Phys.}, 15(9):095003, September 2013.

\bibitem{ramakrishnan_quantum_2014}
Raghunathan Ramakrishnan, Pavlo~O. Dral, Matthias Rupp, and O.~Anatole von
  Lilienfeld.
\newblock Quantum chemistry structures and properties of 134 kilo molecules.
\newblock {\em Sci. Data}, 1(1):140022, August 2014.
\newblock Number: 1 Publisher: Nature Publishing Group.

\bibitem{draxl_nomad_2019}
Claudia Draxl and Matthias Scheffler.
\newblock The {NOMAD} laboratory: from data sharing to artificial intelligence.
\newblock {\em J. Phys. Mater.}, 2(3):036001, May 2019.
\newblock Publisher: IOP Publishing.

\bibitem{stuke_atomic_2020}
Annika Stuke, Christian Kunkel, Dorothea Golze, Milica Todorović, Johannes~T.
  Margraf, Karsten Reuter, Patrick Rinke, and Harald Oberhofer.
\newblock Atomic structures and orbital energies of 61,489 crystal-forming
  organic molecules.
\newblock {\em Sci. Data}, 7(1):58, February 2020.
\newblock Number: 1 Publisher: Nature Publishing Group.

\bibitem{fernandez_machine_2017}
Michael Fernandez, Ante Bilić, and Amanda~S. Barnard.
\newblock Machine learning and genetic algorithm prediction of energy
  differences between electronic calculations of graphene nanoflakes.
\newblock {\em Nanotechnology}, 28(38):38LT03, August 2017.
\newblock Publisher: IOP Publishing.

\bibitem{sun_machine_2017}
Baichuan Sun, Michael Fernandez, and Amanda~S. Barnard.
\newblock Machine {Learning} for {Silver} {Nanoparticle} {Electron} {Transfer}
  {Property} {Prediction}.
\newblock {\em J. Chem. Inf. Model.}, 57(10):2413--2423, October 2017.
\newblock Publisher: American Chemical Society.

\bibitem{barnard_predicting_2019}
A.~S. Barnard and G.~Opletal.
\newblock Predicting structure/property relationships in multi-dimensional
  nanoparticle data using t-distributed stochastic neighbour embedding and
  machine learning.
\newblock {\em Nanoscale}, 11(48):23165--23172, 2019.

\bibitem{barnard_does_2019}
Amanda~S. Barnard, George Opletal, and Shery L.~Y. Chang.
\newblock Does {Twinning} {Impact} {Structure}/{Property} {Relationships} in
  {Diamond} {Nanoparticles}?
\newblock {\em J. Phys. Chem. C}, 123(17):11207--11215, May 2019.

\bibitem{furxhi_machine_2019}
Irini Furxhi, Finbarr Murphy, Martin Mullins, and Craig~A. Poland.
\newblock Machine learning prediction of nanoparticle in vitro toxicity: {A}
  comparative study of classifiers and ensemble-classifiers using the
  {Copeland} {Index}.
\newblock {\em Toxicol. Lett.}, 312:157--166, September 2019.

\bibitem{jparker_machine_2020}
Amanda J. Parker and Amanda S. Barnard.
\newblock Machine learning reveals multiple classes of diamond nanoparticles.
\newblock {\em Nanoscale Horiz.}, 5(10):1394--1399, 2020.
\newblock Publisher: Royal Society of Chemistry.

\bibitem{weber_theoretical_2019}
Fabian Weber, Jian Ren, Tristan Petit, and Annika Bande.
\newblock Theoretical {X}-ray absorption spectroscopy database analysis for
  oxidised {2D} carbon nanomaterials.
\newblock {\em Phys. Chem. Chem. Phys.}, 21(13):6999--7008, 2019.

\bibitem{daly_learning_2020}
Clyde~A. Daly and Rigoberto Hernandez.
\newblock Learning from the {Machine}: {Uncovering} {Sustainable}
  {Nanoparticle} {Design} {Rules}.
\newblock {\em J. Phys. Chem. C}, 124(24):13409--13420, June 2020.
\newblock Publisher: American Chemical Society.

\bibitem{luong_boron-doped_2009}
John H.~T. Luong, Keith~B. Male, and Jeremy~D. Glennon.
\newblock Boron-doped diamond electrode: synthesis, characterization,
  functionalization and analytical applications.
\newblock {\em Analyst}, 134(10):1965, 2009.

\bibitem{denisenko_surface_2010}
A.~Denisenko, A.~Romanyuk, C.~Pietzka, J.~Scharpf, and E.~Kohn.
\newblock Surface structure and surface barrier characteristics of boron-doped
  diamond in electrolytes after {CF4} plasma treatment in {RF}-barrel reactor.
\newblock {\em Diam. Relat. Mater.}, 19(5):423--427, May 2010.

\bibitem{acosta_nitrogen-vacancy_2013}
Victor Acosta and Philip Hemmer.
\newblock Nitrogen-vacancy centers: {Physics} and applications.
\newblock {\em MRS Bull.}, 38(2):127--130, February 2013.

\bibitem{bannwarth_extended_2021}
Christoph Bannwarth, Eike Caldeweyher, Sebastian Ehlert, Andreas Hansen,
  Philipp Pracht, Jakob Seibert, Sebastian Spicher, and Stefan Grimme.
\newblock Extended tight-binding quantum chemistry methods.
\newblock {\em WIREs Comput. Mol. Sci.}, 11(2):e1493, 2021.
\newblock \_eprint:
  https://wires.onlinelibrary.wiley.com/doi/pdf/10.1002/wcms.1493.

\bibitem{bannwarth_gfn2_2019}
Christoph Bannwarth, Sebastian Ehlert, and Stefan Grimme.
\newblock {GFN2}-{xTB}—{An} {Accurate} and {Broadly} {Parametrized}
  {Self}-{Consistent} {Tight}-{Binding} {Quantum} {Chemical} {Method} with
  {Multipole} {Electrostatics} and {Density}-{Dependent} {Dispersion}
  {Contributions}.
\newblock {\em J. Chem. Theory Comput.}, 15(3):1652--1671, March 2019.
\newblock Publisher: American Chemical Society.

\bibitem{neese_orca_2020}
Frank Neese, Frank Wennmohs, Ute Becker, and Christoph Riplinger.
\newblock The {ORCA} quantum chemistry program package.
\newblock {\em J. Chem. Phys.}, 152(22):224108, June 2020.
\newblock Publisher: American Institute of Physics.

\bibitem{adamo_toward_1999}
Carlo Adamo and Vincenzo Barone.
\newblock Toward reliable density functional methods without adjustable
  parameters: {The} {PBE0} model.
\newblock {\em J. Chem. Phys.}, 110(13):6158--6170, April 1999.
\newblock Publisher: American Institute of Physics.

\bibitem{perdew_rationale_1996}
John~P. Perdew, Matthias Ernzerhof, and Kieron Burke.
\newblock Rationale for mixing exact exchange with density functional
  approximations.
\newblock {\em J. Chem. Phys.}, 105(22):9982--9985, December 1996.
\newblock Publisher: American Institute of Physics.

\bibitem{schafer_fully_1992}
Ansgar Schäfer, Hans Horn, and Reinhart Ahlrichs.
\newblock Fully optimized contracted {Gaussian} basis sets for atoms {Li} to
  {Kr}.
\newblock {\em J. Chem. Phys.}, 97(4):2571--2577, August 1992.

\bibitem{weigend_balanced_2005}
Florian Weigend and Reinhart Ahlrichs.
\newblock Balanced basis sets of split valence, triple zeta valence and
  quadruple zeta valence quality for {H} to {Rn}: {Design} and assessment of
  accuracy.
\newblock {\em Phys. Chem. Chem. Phys.}, 7(18):3297--3305, August 2005.

\bibitem{grimme_effect_2011}
Stefan Grimme, Stephan Ehrlich, and Lars Goerigk.
\newblock Effect of the damping function in dispersion corrected density
  functional theory.
\newblock {\em J. Comput. Chem.}, 32(7):1456--1465, 2011.

\bibitem{neese_efficient_2009}
Frank Neese, Frank Wennmohs, Andreas Hansen, and Ute Becker.
\newblock Efficient, approximate and parallel {Hartree}–{Fock} and hybrid
  {DFT} calculations. {A} ‘chain-of-spheres’ algorithm for the
  {Hartree}–{Fock} exchange.
\newblock {\em Chem. Phys.}, 356(1):98--109, February 2009.

\bibitem{weigend_accurate_2006}
Florian Weigend.
\newblock Accurate {Coulomb}-fitting basis sets for {H} to {Rn}.
\newblock {\em Phys. Chem. Chem. Phys.}, 8(9):1057--1065, February 2006.

\bibitem{unke_machine_2021}
Oliver~T. Unke, Stefan Chmiela, Huziel~E. Sauceda, Michael Gastegger, Igor
  Poltavsky, Kristof~T. Schütt, Alexandre Tkatchenko, and Klaus-Robert
  Müller.
\newblock Machine {Learning} {Force} {Fields}.
\newblock {\em Chem. Rev.}, 121(16):10142--10186, August 2021.
\newblock Publisher: American Chemical Society.

\bibitem{cho2014properties}
Kyunghyun Cho, Bart Van~Merri{\"e}nboer, Dzmitry Bahdanau, and Yoshua Bengio.
\newblock On the properties of neural machine translation: Encoder-decoder
  approaches.
\newblock {\em arXiv preprint arXiv:1409.1259}, 2014.

\bibitem{vinyals2015order}
Oriol Vinyals, Samy Bengio, and Manjunath Kudlur.
\newblock Order matters: Sequence to sequence for sets.
\newblock {\em arXiv preprint arXiv:1511.06391}, 2015.

\bibitem{schutt_schnet_2018}
K.~T. Schütt, H.~E. Sauceda, P.-J. Kindermans, A.~Tkatchenko, and K.-R.
  Müller.
\newblock {SchNet} – {A} deep learning architecture for molecules and
  materials.
\newblock {\em J. Chem. Phys.}, 148(24):241722, March 2018.
\newblock Publisher: American Institute of Physics.

\bibitem{schutt_schnetpack_2019}
K.~T. Schütt, P.~Kessel, M.~Gastegger, K.~A. Nicoli, A.~Tkatchenko, and K.-R.
  Müller.
\newblock {SchNetPack}: {A} {Deep} {Learning} {Toolbox} {For} {Atomistic}
  {Systems}.
\newblock {\em J. Chem. Theory Comput.}, 15(1):448--455, January 2019.
\newblock Publisher: American Chemical Society.

\bibitem{klicpera2020directional}
Johannes Klicpera, Janek Gro{\ss}, and Stephan G{\"u}nnemann.
\newblock Directional message passing for molecular graphs.
\newblock {\em arXiv preprint arXiv:2003.03123}, 2020.

\bibitem{klicpera2020fast}
Johannes Klicpera, Shankari Giri, Johannes~T Margraf, and Stephan
  G{\"u}nnemann.
\newblock Fast and uncertainty-aware directional message passing for
  non-equilibrium molecules.
\newblock {\em arXiv preprint arXiv:2011.14115}, 2020.

\bibitem{abdi_principal_2010}
Hervé Abdi and Lynne~J. Williams.
\newblock Principal component analysis.
\newblock {\em Wiley Interdiscip. Rev. Comput. Stat.}, 2(4):433--459, 2010.

\bibitem{casier_using_2020}
Bastien Casier, Stéphane Carniato, Tsveta Miteva, Nathalie Capron, and Nicolas
  Sisourat.
\newblock Using principal component analysis for neural network
  high-dimensional potential energy surface.
\newblock {\em J. Chem. Phys.}, 152(23):234103, 2020.

\bibitem{darby_compressing_2022}
James~P. Darby, James~R. Kermode, and Gábor Csányi.
\newblock Compressing local atomic neighbourhood descriptors.
\newblock {\em npj Comput. Mater.}, 8(1):1--13, August 2022.
\newblock Number: 1 Publisher: Nature Publishing Group.

\bibitem{musil2018librascal}
F{\'e}lix Musil, Max Veit, Till Junge, Markus Stricker, Alexander Goscinki,
  Guillaume Fraux, and Michele Ceriotti.
\newblock Librascal.
\newblock {\em GitHub, https://github. com/cosmo-epfl/librascal}, 2018.

\bibitem{scikit-learn}
F.~Pedregosa, G.~Varoquaux, A.~Gramfort, V.~Michel, B.~Thirion, O.~Grisel,
  M.~Blondel, P.~Prettenhofer, R.~Weiss, V.~Dubourg, J.~Vanderplas, A.~Passos,
  D.~Cournapeau, M.~Brucher, M.~Perrot, and E.~Duchesnay.
\newblock Scikit-learn: Machine learning in {P}ython.
\newblock {\em J. Mach. Learn Res.}, 12:2825--2830, 2011.

\bibitem{DBLP:journals/corr/abs-1912-01703}
Adam Paszke, Sam Gross, Francisco Massa, Adam Lerer, James Bradbury, Gregory
  Chanan, Trevor Killeen, Zeming Lin, Natalia Gimelshein, Luca Antiga, Alban
  Desmaison, Andreas K{\"{o}}pf, Edward~Z. Yang, Zach DeVito, Martin Raison,
  Alykhan Tejani, Sasank Chilamkurthy, Benoit Steiner, Lu~Fang, Junjie Bai, and
  Soumith Chintala.
\newblock Pytorch: An imperative style, high-performance deep learning library.
\newblock {\em arXiv}, arXiv:1912.01703, 2019.

\bibitem{Fey/Lenssen/2019}
Matthias Fey and Jan~E. Lenssen.
\newblock Fast graph representation learning with {PyTorch Geometric}.
\newblock In {\em ICLR Workshop on Representation Learning on Graphs and
  Manifolds}, 2019.

\bibitem{dscribe}
Lauri Himanen, Marc O.~J. Jäger, Eiaki~V. Morooka, Filippo Federici~Canova,
  Yashasvi~S. Ranawat, David~Z. Gao, Patrick Rinke, and Adam~S. Foster.
\newblock {DScribe: Library of descriptors for machine learning in materials
  science}.
\newblock {\em Comput. Phys. Commun.}, 247:106949, 2020.

\bibitem{Hjorth_Larsen_2017}
Ask~Hjorth Larsen, Jens~J{\o}rgen Mortensen, Jakob Blomqvist, Ivano~E Castelli,
  Rune Christensen, Marcin Du{\l}ak, Jesper Friis, Michael~N Groves, Bj{\o}rk
  Hammer, Cory Hargus, Eric~D Hermes, Paul~C Jennings, Peter~Bjerre Jensen,
  James Kermode, John~R Kitchin, Esben~Leonhard Kolsbjerg, Joseph Kubal,
  Kristen Kaasbjerg, Steen Lysgaard, J{\'{o}}n~Bergmann Maronsson, Tristan
  Maxson, Thomas Olsen, Lars Pastewka, Andrew Peterson, Carsten Rostgaard,
  Jakob Schi{\o}tz, Ole Schütt, Mikkel Strange, Kristian~S Thygesen, Tejs
  Vegge, Lasse Vilhelmsen, Michael Walter, Zhenhua Zeng, and Karsten~W
  Jacobsen.
\newblock The atomic simulation environment{\textemdash}a python library for
  working with atoms.
\newblock {\em J. Phys. Condens. Mat.}, 29(27):273002, 2017.

\bibitem{liu_functionalization_2004}
Yu~Liu, Zhenning Gu, John~L. Margrave, and Valery~N. Khabashesku.
\newblock Functionalization of {Nanoscale} {Diamond} {Powder}: {Fluoro}-,
  {Alkyl}-, {Amino}-, and {Amino} {Acid}-{Nanodiamond} {Derivatives}.
\newblock {\em Chem. Mater.}, 16(20):3924--3930, October 2004.

\bibitem{panich_structure_2010}
Alexander~M. Panich, Hans-Martin Vieth, Alexander~I. Shames, Natalya Froumin,
  Eiji Ôsawa, and Akifumi Yao.
\newblock Structure and {Bonding} in {Fluorinated} {Nanodiamond}.
\newblock {\em J. Phys. Chem. C}, 114(2):774--782, January 2010.

\bibitem{shenderova_hydroxylated_2011}
O.~Shenderova, A.~M. Panich, S.~Moseenkov, S.~C. Hens, V.~Kuznetsov, and H.-M.
  Vieth.
\newblock Hydroxylated {Detonation} {Nanodiamond}: {FTIR}, {XPS}, and {NMR}
  {Studies}.
\newblock {\em J. Phys. Chem. C}, 115(39):19005--19011, October 2011.
\newblock Publisher: American Chemical Society.

\bibitem{zhu_amino-terminated_2016}
Di~Zhu, Jason~A. Bandy, Shuo Li, and Robert~J. Hamers.
\newblock Amino-terminated diamond surfaces: {Photoelectron} emission and
  photocatalytic properties.
\newblock {\em Surf. Sci.}, 650:295--301, August 2016.

\bibitem{kajihara_nitrogen_1991}
S.~A. Kajihara, A.~Antonelli, J.~Bernholc, and R.~Car.
\newblock Nitrogen and potential n-type dopants in diamond.
\newblock {\em Phys. Rev. Lett.}, 66(15):2010--2013, April 1991.
\newblock Publisher: American Physical Society.

\bibitem{cui_si-doped_2013}
Yu-xiao Cui, Jian-guo Zhang, Fang-hong Sun, and Zhi-ming Zhang.
\newblock Si-doped diamond films prepared by chemical vapour deposition.
\newblock {\em T. Nonferr. Metal. Soc.}, 23(10):2962--2970, October 2013.

\bibitem{wang_influences_2020}
Xinchang Wang, Yu~Qiao, Baocai Zhang, and Fanghong Sun.
\newblock Influences of {Si} dopant on geometry and energetic stability of
  terminated diamond (111)-1 × 1 surfaces.
\newblock {\em Diam. Relat. Mater.}, 109:108014, November 2020.

\bibitem{wei_effect_2020}
Xubing Wei, Lin Chen, Minglan Zhang, Zhibin Lu, and Guangan Zhang.
\newblock Effect of dopants ({F}, {Si}) material on the structure and
  properties of hydrogenated {DLC} film by plane cathode {PECVD}.
\newblock {\em Diam. Relat. Mater.}, 110:108102, December 2020.

\bibitem{grotjohn_heavy_2014}
T.~A. Grotjohn, D.~T. Tran, M.~K. Yaran, S.~N. Demlow, and T.~Schuelke.
\newblock Heavy phosphorus doping by epitaxial growth on the (111) diamond
  surface.
\newblock {\em Diam. Relat. Mater.}, 44:129--133, April 2014.

\bibitem{kato_heavily_2016}
Hiromitsu Kato, Daisuke Takeuchi, Masahiko Ogura, Takatoshi Yamada, Mitsuhiro
  Kataoka, Yuji Kimura, Susumu Sobue, Christoph~E. Nebel, and Satoshi Yamasaki.
\newblock Heavily phosphorus-doped nano-crystalline diamond electrode for
  thermionic emission application.
\newblock {\em Diam. Relat. Mater.}, 63:165--168, March 2016.

\bibitem{alfieri_phosphorus-related_2018}
Giovanni Alfieri, Lukas Kranz, and Andrei Mihaila.
\newblock Phosphorus-{Related} {Complexes} and {Shallow} {Doping} in {Diamond}.
\newblock {\em Phys. Status Solidi Rapid Res. Lett.}, 12(4):1700409, 2018.
\newblock \_eprint:
  https://onlinelibrary.wiley.com/doi/pdf/10.1002/pssr.201700409.

\bibitem{ristein_surface_2006}
Jürgen Ristein.
\newblock Surface science of diamond: {Familiar} and amazing.
\newblock {\em Surf. Sci.}, 600(18):3677--3689, September 2006.

\bibitem{buchner_early_2021}
Franziska Buchner, Thorren Kirschbaum, Amélie Venerosy, Hugues Girard,
  Jean-Charles Arnault, Benjamin Kiendl, Anke Krueger, Karin Larsson, Annika
  Bande, Tristan Petit, and Christoph Merschjann.
\newblock Early dynamics of the emission of solvated electrons from
  nanodiamonds in water.
\newblock {\em chemarXiv preprint chemarXiv:10.26434}, December 2021.

\bibitem{christianson_mechanism_2014}
Jeffrey~R. Christianson, Di~Zhu, Robert~J. Hamers, and J.~R. Schmidt.
\newblock Mechanism of {N2} {Reduction} to {NH3} by {Aqueous} {Solvated}
  {Electrons}.
\newblock {\em J. Phys. Chem. B}, 118(1):195--203, January 2014.
\newblock Publisher: American Chemical Society.

\bibitem{ambrosio_absolute_2018}
Francesco Ambrosio, Zhendong Guo, and Alfredo Pasquarello.
\newblock Absolute {Energy} {Levels} of {Liquid} {Water}.
\newblock {\em J. Phys. Chem. Lett.}, 9(12):3212--3216, June 2018.
\newblock Publisher: American Chemical Society.

\bibitem{han_surface-bound_2017}
Peng Han, Denis Antonov, Jörg Wrachtrup, and Gabriel Bester.
\newblock Surface-bound states in nanodiamonds.
\newblock {\em Phys. Rev. B}, 95(19):195428, May 2017.

\bibitem{faber_prediction_2017}
Felix~A. Faber, Luke Hutchison, Bing Huang, Justin Gilmer, Samuel~S.
  Schoenholz, George~E. Dahl, Oriol Vinyals, Steven Kearnes, Patrick~F. Riley,
  and O.~Anatole von Lilienfeld.
\newblock Prediction {Errors} of {Molecular} {Machine} {Learning} {Models}
  {Lower} than {Hybrid} {DFT} {Error}.
\newblock {\em J. Chem. Theory Comput.}, 13(11):5255--5264, November 2017.
\newblock Publisher: American Chemical Society.

\bibitem{yaghoobi_machine_2022}
Mostafa Yaghoobi and Mojtaba Alaei.
\newblock Machine learning for compositional disorder: {A} comparison between
  different descriptors and machine learning frameworks.
\newblock {\em Comp. Mater. Sci.}, 207:111284, May 2022.

\bibitem{liu_transferable_2021}
Ziteng Liu, Liqiang Lin, Qingqing Jia, Zheng Cheng, Yanyan Jiang, Yanwen Guo,
  and Jing Ma.
\newblock Transferable {Multilevel} {Attention} {Neural} {Network} for
  {Accurate} {Prediction} of {Quantum} {Chemistry} {Properties} via {Multitask}
  {Learning}.
\newblock {\em J. Chem. Inf. Model.}, 61(3):1066--1082, March 2021.
\newblock Publisher: American Chemical Society.

\bibitem{bolker_quantum_2013}
Asaf Bolker, Cecile Saguy, Moshe Tordjman, and Rafi Kalish.
\newblock Quantum confinement and {Coulomb} blockade in isolated nanodiamond
  crystallites.
\newblock {\em Phys. Rev. B}, 88(3):035442, July 2013.
\newblock Publisher: American Physical Society.

\bibitem{pearson1903laws}
Karl Pearson and Alice Lee.
\newblock On the laws of inheritance in man: I. inheritance of physical
  characters.
\newblock {\em Biometrika}, 2(4):357--462, 1903.

\bibitem{bennett_curta_2020}
Loris Bennett, Bernd Melchers, and Boris Proppe.
\newblock Curta: {A} {General}-purpose {High}-{Performance} {Computer} at
  {ZEDAT}, {Freie} {Universität} {Berlin}.
\newblock 2020.
\newblock Accepted: 2020-03-19T11:06:50Z.

\end{thebibliography}


\begin{thebibliography}{10}

\bibitem{bannwarth_extended_2021}
Christoph Bannwarth, Eike Caldeweyher, Sebastian Ehlert, Andreas Hansen,
  Philipp Pracht, Jakob Seibert, Sebastian Spicher, and Stefan Grimme.
\newblock Extended tight-binding quantum chemistry methods.
\newblock {\em WIREs Comput. Mol. Sci.}, 11(2):e1493, 2021.
\newblock \_eprint:
  https://wires.onlinelibrary.wiley.com/doi/pdf/10.1002/wcms.1493.

\bibitem{lopez-carballeira_ab_2020}
Diego López-Carballeira and Tomáš Polcar.
\newblock Ab initio description of nanodiamonds: {A} {DFT} and {TDDFT}
  benchmark.
\newblock {\em Diam. Relat. Mater.}, 108:107959, October 2020.

\bibitem{perdew_generalized_1996}
John~P. Perdew, Kieron Burke, and Matthias Ernzerhof.
\newblock Generalized {Gradient} {Approximation} {Made} {Simple}.
\newblock {\em Phys. Rev. Lett.}, 77(18):3865--3868, October 1996.
\newblock Publisher: American Physical Society.

\bibitem{grimme_effect_2011}
Stefan Grimme, Stephan Ehrlich, and Lars Goerigk.
\newblock Effect of the damping function in dispersion corrected density
  functional theory.
\newblock {\em J. Comput. Chem.}, 32(7):1456--1465, 2011.

\bibitem{schafer_fully_1992}
Ansgar Schäfer, Hans Horn, and Reinhart Ahlrichs.
\newblock Fully optimized contracted {Gaussian} basis sets for atoms {Li} to
  {Kr}.
\newblock {\em J. Chem. Phys.}, 97(4):2571--2577, August 1992.

\bibitem{weigend_balanced_2005}
Florian Weigend and Reinhart Ahlrichs.
\newblock Balanced basis sets of split valence, triple zeta valence and
  quadruple zeta valence quality for {H} to {Rn}: {Design} and assessment of
  accuracy.
\newblock {\em Phys. Chem. Chem. Phys.}, 7(18):3297--3305, August 2005.

\bibitem{adamo_toward_1999}
Carlo Adamo and Vincenzo Barone.
\newblock Toward reliable density functional methods without adjustable
  parameters: {The} {PBE0} model.
\newblock {\em J. Chem. Phys.}, 110(13):6158--6170, April 1999.
\newblock Publisher: American Institute of Physics.

\bibitem{perdew_rationale_1996}
John~P. Perdew, Matthias Ernzerhof, and Kieron Burke.
\newblock Rationale for mixing exact exchange with density functional
  approximations.
\newblock {\em J. Chem. Phys.}, 105(22):9982--9985, December 1996.
\newblock Publisher: American Institute of Physics.

\bibitem{stewart_optimization_2013}
James J.~P. Stewart.
\newblock Optimization of parameters for semiempirical methods {VI}: more
  modifications to the {NDDO} approximations and re-optimization of parameters.
\newblock {\em J Mol Model}, 19(1):1--32, January 2013.
\newblock Company: Springer Distributor: Springer Institution: Springer Label:
  Springer Number: 1 Publisher: Springer-Verlag.

\bibitem{frisch2gaussian}
MJ~Frisch, GW~Trucks, HB~Schlegel, GE~Scuseria, MA~Robb, JR~Cheeseman,
  G~Scalmani, V~Barone, GA~Petersson, H~Nakatsuji, et~al.
\newblock Gaussian 16 revision a. 03. 2016; gaussian inc.
\newblock {\em Wallingford CT}, 2(3):4.

\bibitem{willey_molecular_2005}
T.~M. Willey, C.~Bostedt, T.~van Buuren, J.~E. Dahl, S.~G. Liu, R.~M.~K.
  Carlson, L.~J. Terminello, and T.~Möller.
\newblock Molecular {Limits} to the {Quantum} {Confinement} {Model} in
  {Diamond} {Clusters}.
\newblock {\em Phys. Rev. Lett.}, 95(11):113401, September 2005.
\newblock Publisher: American Physical Society.

\bibitem{teunissen_tuning_2017}
Jos~L. Teunissen, Frank De~Proft, and Freija De~Vleeschouwer.
\newblock Tuning the {HOMO}–{LUMO} {Energy} {Gap} of {Small} {Diamondoids}
  {Using} {Inverse} {Molecular} {Design}.
\newblock {\em J. Chem. Theory Comput.}, 13(3):1351--1365, March 2017.

\bibitem{musil2018librascal}
F{\'e}lix Musil, Max Veit, Till Junge, Markus Stricker, Alexander Goscinki,
  Guillaume Fraux, and Michele Ceriotti.
\newblock Librascal.
\newblock {\em GitHub, https://github. com/cosmo-epfl/librascal}, 2018.

\bibitem{Fey/Lenssen/2019}
Matthias Fey and Jan~E. Lenssen.
\newblock Fast graph representation learning with {PyTorch Geometric}.
\newblock In {\em ICLR Workshop on Representation Learning on Graphs and
  Manifolds}, 2019.

\bibitem{stuke_atomic_2020}
Annika Stuke, Christian Kunkel, Dorothea Golze, Milica Todorović, Johannes~T.
  Margraf, Karsten Reuter, Patrick Rinke, and Harald Oberhofer.
\newblock Atomic structures and orbital energies of 61,489 crystal-forming
  organic molecules.
\newblock {\em Sci. Data}, 7(1):58, February 2020.
\newblock Number: 1 Publisher: Nature Publishing Group.

\bibitem{rahaman_deep_2020}
Obaidur Rahaman and Alessio Gagliardi.
\newblock Deep {Learning} {Total} {Energies} and {Orbital} {Energies} of
  {Large} {Organic} {Molecules} {Using} {Hybridization} of {Molecular}
  {Fingerprints}.
\newblock {\em J. Chem. Inf. Model.}, 60(12):5971--5983, December 2020.
\newblock Publisher: American Chemical Society.

\bibitem{caro_optimizing_2019}
Miguel~A. Caro.
\newblock Optimizing many-body atomic descriptors for enhanced computational
  performance of machine learning based interatomic potentials.
\newblock {\em Phys. Rev. B}, 100(2):024112, July 2019.
\newblock Publisher: American Physical Society.

\bibitem{schutt_equivariant_2021}
Kristof Schütt, Oliver Unke, and Michael Gastegger.
\newblock Equivariant message passing for the prediction of tensorial
  properties and molecular spectra.
\newblock In {\em Proceedings of the 38th {International} {Conference} on
  {Machine} {Learning}}, pages 9377--9388. PMLR, July 2021.
\newblock ISSN: 2640-3498.

\bibitem{ross_incremental_2008}
David~A. Ross, Jongwoo Lim, Ruei-Sung Lin, and Ming-Hsuan Yang.
\newblock Incremental {Learning} for {Robust} {Visual} {Tracking}.
\newblock {\em Int J Comput Vis}, 77(1):125--141, May 2008.

\end{thebibliography}
\bibliographystyle{unsrt}

\end{document}


\newcommand{\ba}{\begin{eqnarray}}
\newcommand{\ea}{\end{eqnarray}}
\newcommand{\bc}{\begin{center}}
\newcommand{\ec}{\end{center}}
\newcommand{\be}{\begin{equation}}
\newcommand{\ee}{\end{equation}}
\newcommand{\EA}{\emph{et al.}\xspace}
\newcommand{\ra}{\rightarrow}
\newcommand{\bi}{\begin{itemize}}
\newcommand{\ei}{\end{itemize}}
\newcommand{\rb}{{\bf r}}

\centering \huge Supporting Information for Machine Learning Frontier Orbital Energies of Nanodiamonds \\ ~\\
\large Thorren Kirschbaum\textsuperscript{1,2}, Börries von Seggern\textsuperscript{1,3}, Joachim Dzubiella\textsuperscript{1,4}, Annika Bande\textsuperscript{1$\ast$}, Frank Noé\textsuperscript{5,2,6,7$\ast$} \\ ~\\
\raggedright \normalsize \textsuperscript{1}Helmholtz-Zentrum Berlin für Materialien und Energie GmbH, Hahn-Meitner-Platz 1, 14109 Berlin, Germany\\
\textsuperscript{2}Department of Mathematics and Computer Science, Freie Universität Berlin, Arnimallee 12, 14195 Berlin, Germany\\
\textsuperscript{3}Department of Biology, Chemistry and Pharmacy, Freie Universität Berlin, Arnimallee 22, 14195 Berlin, Germany\\
\textsuperscript{4}Institute of Physics, Albert-Ludwigs-Universität Freiburg, Hermann-Herder-Straße 3, 79104 Freiburg im Breisgau, Germany\\
\textsuperscript{5}Microsoft Research AI4Science, Karl-Liebknecht Str. 32, 10178 Berlin, Germany\\
\textsuperscript{6}Department of Physics, Freie Universität Berlin, Arnimallee 12, 14195 Berlin, Germany\\
\textsuperscript{7}Department of Chemistry, Rice University, 6100 Main St, Houston, TX 77005, United States\\
 ~\\
\textsuperscript{$\ast$}Corresponding Author: annika.bande@helmholtz-berlin.de\\
\textsuperscript{$\ast$}Corresponding Author: franknoe@microsoft.com

\clearpage

\section{ND5k Base Structures}\label{SecBS}

\begin{justify}

Figure \ref{fig:BS} displays the 17 undoped, H-terminated base structures used to compile the ND5k dataset. In the dataset file and in the section below, the diamondoid structures will be abbreviated as follows: Adamantane (ad), Diamantane (di), Triamantane (tri), [123]Tetramantane (123tet), [121]Tetramantane (121tet), [1(2)3]Tetramantane (1-2-3tet), [1212]Pentamantane (1212pent), [1(23)4]Pentamantane (1234pent), [12312]Hexamantane (12312hex).

\begin{figure}[h]
 \makeatletter 
 \renewcommand{\thefigure}{S\@arabic\c@figure}
 \makeatother
 \centering
 \includegraphics[width = 1.0\textwidth]{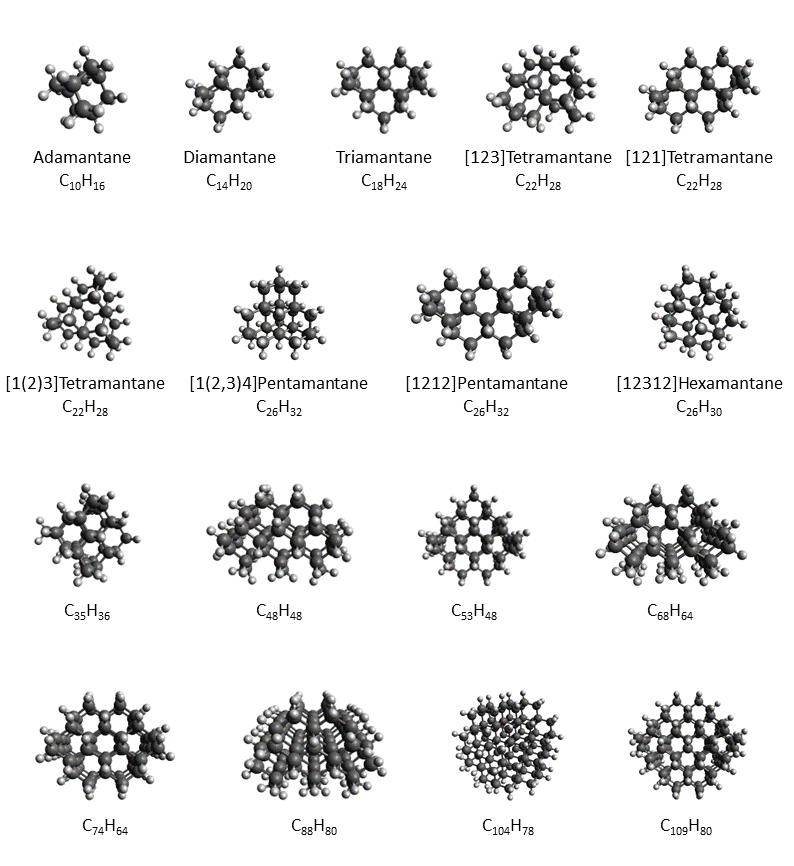}
    \caption{Base structures of the ND5k dataset.}
    \label{fig:BS}
\end{figure} 

\clearpage

\section{DFTB Geometry Optimization of Nanodiamond Structures}\label{SecDFTB}

The DFTB geometry optimization algorithm\cite{bannwarth_extended_2021} was tested against standard DFT optimization at PBE-D3/SVP level of theory.\cite{lopez-carballeira_ab_2020, perdew_generalized_1996, grimme_effect_2011, schafer_fully_1992, weigend_balanced_2005} For 11 randomly chosen NDs, structures were optimized with each method, and subsequent single point calculations were performed on PBE0-D3/SVP level of theory.\cite{adamo_toward_1999, perdew_rationale_1996, grimme_effect_2011, schafer_fully_1992, weigend_balanced_2005} For the PBE0 HOMO and LUMO values, we found the following deviations (DFTB structures benchmarked against PBE structures, MAD is mean absolute deviation): 

HOMO: MAD: 0.230 eV, max. abs. dev.: 0.504 eV, min. abs. dev.: 0.027 eV 

LUMO: MAD: 0.195 eV, max. abs. dev.: 0.491 eV, min. abs. dev.: 0.008 eV 

The deviations of DFTB-optimized structures benchmarked against PBE-optimized structures are around 0.2 eV MAD. When looking at the benchmark data in more detail, it is obvious that the deviations obtained for H- and F-terminated NDs (MADs of ca. 0.13 eV) are much smaller than for OH- and NH2-terminated NDs (MADs of ca. 0.36 eV). The latter NDs have many rotational degrees of freedom in their surface moieties and large H-bond networks on their surfaces, thus, we cannot expect the geometry optimization algorithms to find the global minimum energy conformation of the respective structures. Apparently, DFTB and PBE optimization yield different, but probably both still reliable (local) minimum energy structures. Therefore, their PBE0-computed frontier orbital energies differ more than in the case of single-atom (H or F) terminated NDs. The different structures obtained from DFTB and PBE optimization for two NDs are depicted in figures \ref{fig:B-B-OH} and \ref{fig:Si-P-OH}, corresponding to the structures with the largest frontier energy orbital differences in the benchmark set. The large deviations obtained from DFTB-optimized structures involving OH- and NH$_2$-termination benchmarked against PBE-optimized structures likely do not result from shortcomings of the DFTB optimization, but from the fact that both optimization algorithms end up in different local energy minimum structures.

PBE0-computed optical gaps of diamondoids benchmarked against experiments have an MAD of around 1 eV,\cite{lopez-carballeira_ab_2020} which is much higher than the error obtained by using DFTB instead of PBE geometry optimization. The massive computational speedup and the acceptably low error clearly justifies the use of the DFTB method for geometry optimization. We also tested the semiempirical PM7 method\cite{stewart_optimization_2013} as implemented in Gaussian (PM7MOPAC)\cite{frisch2gaussian} for fast structure optimizations, but found the DFTB algorithm to perform better overall, especially being faster and more stable than PM7.

\begin{figure}[h]
 \makeatletter 
 \renewcommand{\thefigure}{S\@arabic\c@figure}
 \makeatother
 \raggedright
 \includegraphics[width = 0.65\textwidth]{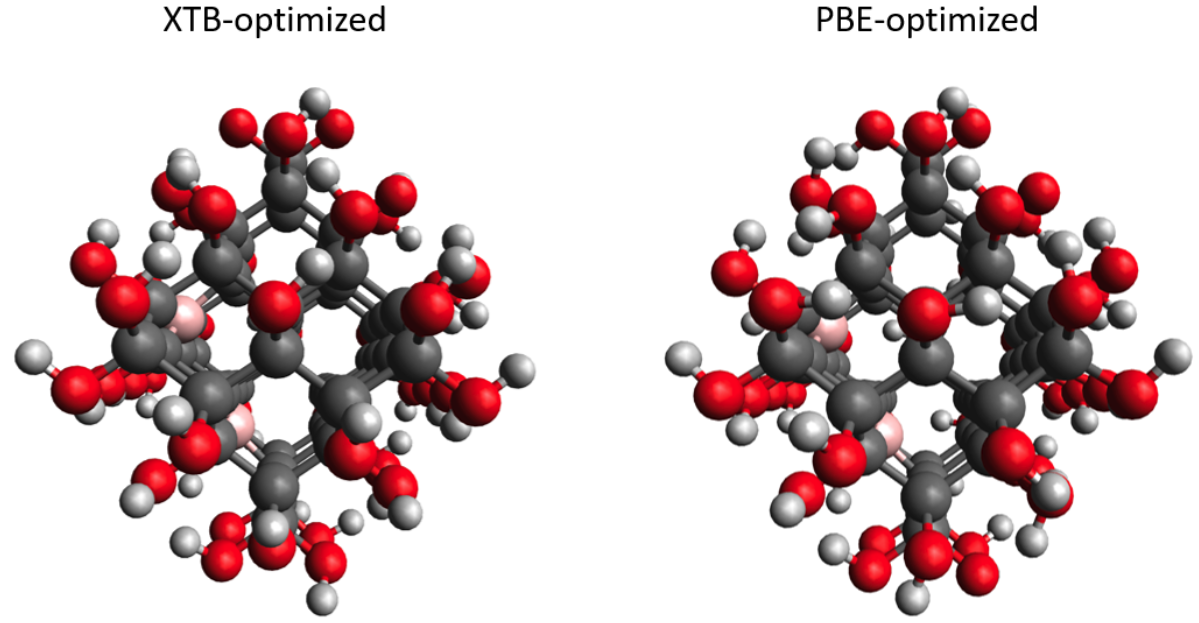}
    \caption{Structure of a B-B doped, OH-terminated ND (base structure C$_{53}$H$_{48}$), optimized by DFTB (left) and PBE (right) shown from the same perspective. The H-bond network formed by the hydroxyl groups is different after both structure optimizations. Color: H (grey), B (rose), C (black), O (red).}
    \label{fig:B-B-OH}
\end{figure}

\begin{figure}[h]
 \makeatletter 
 \renewcommand{\thefigure}{S\@arabic\c@figure}
 \makeatother
 \raggedright
 \medskip
 \includegraphics[width = 0.65\textwidth]{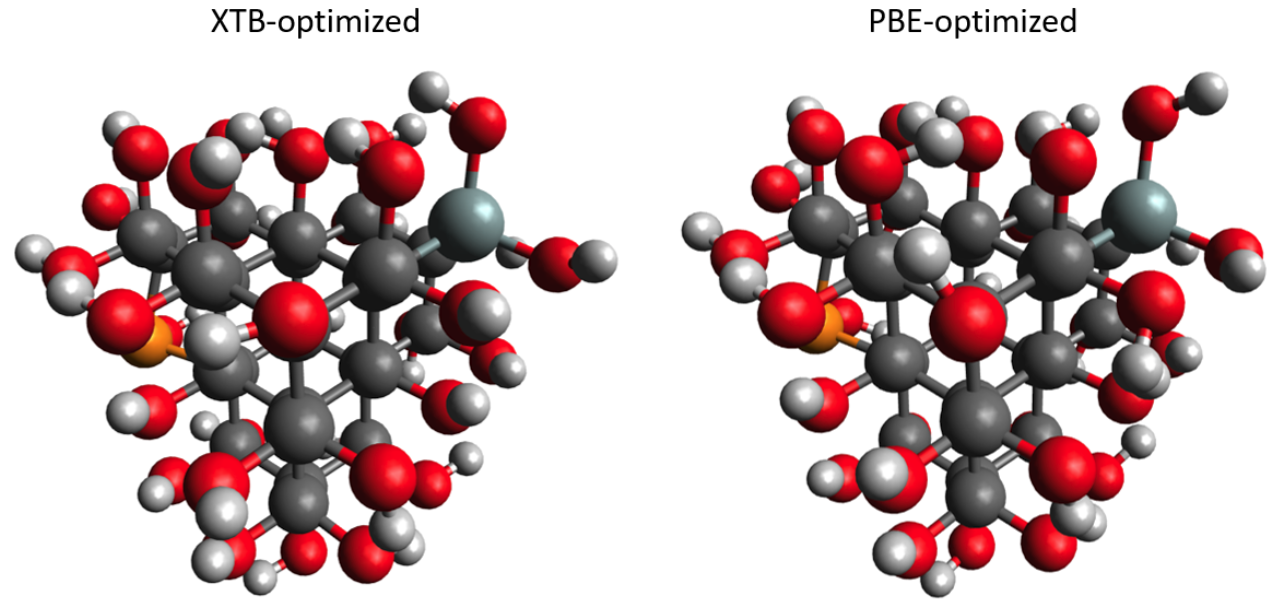}
    \caption{Structure of a Si-P doped, OH-terminated ND (base structure [1(2,3)4]pentamantane), optimized by DFTB (left) and PBE (right) shown from the same perspective. The H-bond network formed by the hydroxyl groups is different after both structure optimizations. Color: H (grey), C (black), O (red), Si (green), P (orange).}
    \label{fig:Si-P-OH}
\end{figure} 

~\\

\section{Frontier Orbitals of P-doped Nanoadiamonds for Photocatalysis}\label{SecMOs}

Here, we display the structures and frontier orbitals of ten of the nanodiamonds that were filtered from the ND5k dataset as candidate structures for use in photocatalysis (see table 1 in the main document). Figures \ref{fig:MOs1} and \ref{fig:MOs2} contain plots of the the structure, HOMO and LUMO shapes for ten P-(co-)doped nanodiamonds.

\begin{figure}[h]
 \makeatletter 
 \renewcommand{\thefigure}{S\@arabic\c@figure}
 \makeatother
 \raggedright
 \includegraphics[width = 0.9\textwidth]{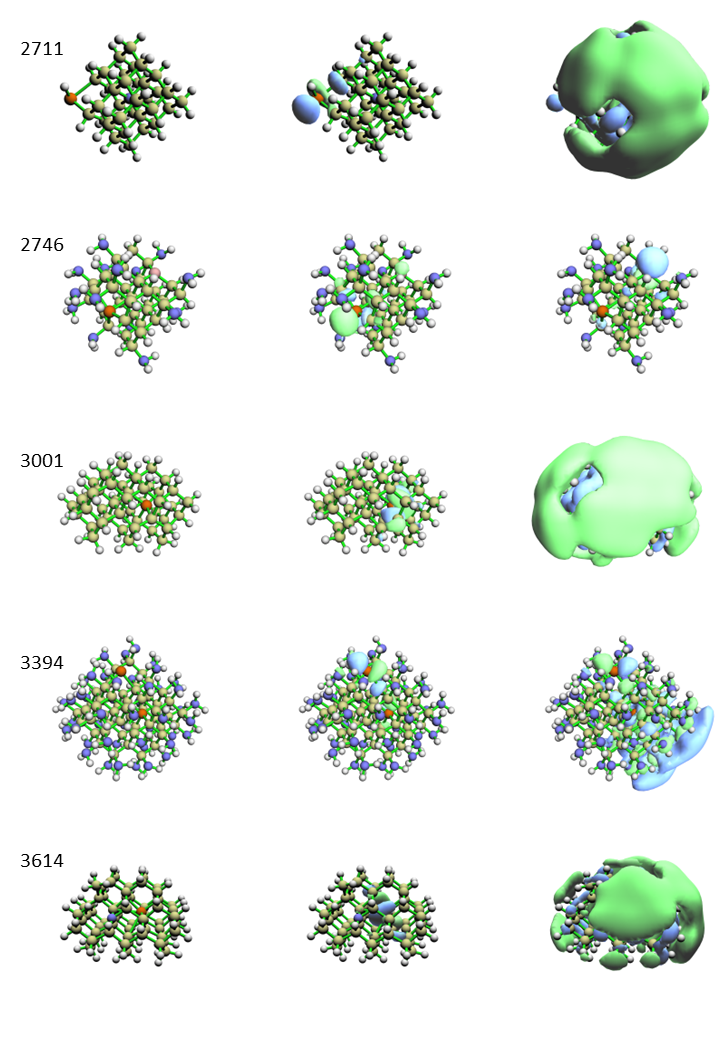}
    \caption{From left to right: ND5k index (cf table 1 in main), structure, HOMO and LUMO contour plot. Color: H (grey), B (rose), C (yellow), N (violet), Si (light brown), P (orange). Isovalues of the contour plots are $\pm$0.05, except fot the diffuse LUMOs of nanodiamonds 2711, 3001, 3394, 3614, where the isovalue is $\pm$ 0.01.}
    \label{fig:MOs1}
\end{figure}

\begin{figure}[h]
 \makeatletter 
 \renewcommand{\thefigure}{S\@arabic\c@figure}
 \makeatother
 \raggedright
 \includegraphics[width = 0.9\textwidth]{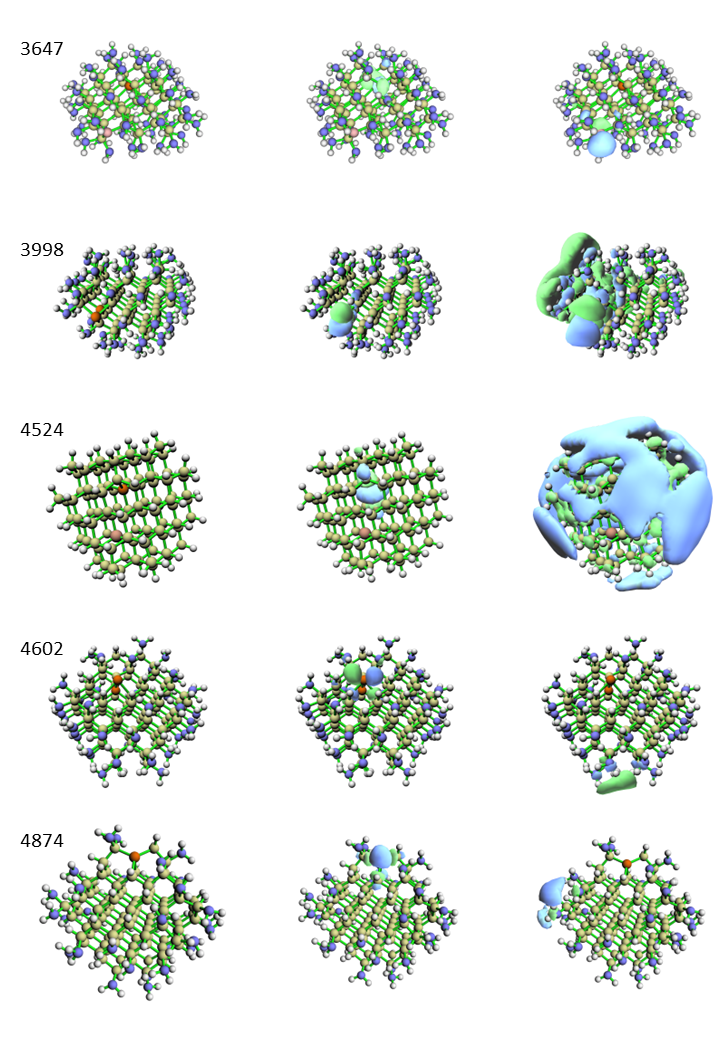}
    \caption{From left to right: ND5k index (cf table 1 in main), structure, HOMO and LUMO contour plot. Color: H (grey), B (rose), C (yellow), N (violet), Si (light brown), P (orange). Isovalues of the contour plots are $\pm$0.05, except fot the diffuse LUMOs of nanodiamonds 3998 and 4524, where the isovalue is $\pm$ 0.01.}
    \label{fig:MOs2}
\end{figure} 

\clearpage

\section{Trends in the ND5k Dataset}\label{SecTrends}

In the following tables S1 to S6 we summarize the ND configurations of the ND5k structures with the highest and lowest HOMO, LUMO and gap energies, respectively. For the cases of undoped and singly doped structures, the "dopant" atom C represents no doping.

\begin{table}[h]
 \makeatletter 
 \renewcommand{\thetable}{S\@arabic\c@table}
 \makeatother
\label{tab:LowHOMO}
\caption{Configurations of the ten ND5k structures with the lowest HOMO energies: ND5k index, base structure (ND), surface species, dopants (D1, D2), HOMO, LUMO and gap energies (all in eV).}
\medskip
\begin{tabular}{llllllll}
Index & ND & Surface & D1 & D2 & E(HOMO) & E(LUMO) & E(gap)   \\
\hline
223      & ad       & F       & C        & C        & $-$10.5967 & $-$1.3038 & 9.293 \\
518      & di       & F       & C        & C        & $-$10.5947 & $-$1.4285 & 9.166 \\
1136     & 1-2-3tet & F       & C        & Si       & $-$10.4318 & $-$2.1995 & 8.232 \\
819      & tri      & F       & C        & C        & $-$10.3975 & $-$1.3823 & 9.015 \\
2324     & 1234pent & F       & C        & C        & $-$10.3749 & $-$0.9683 & 9.407 \\
1120     & 1-2-3tet & F       & C        & C        & $-$10.3486 & $-$1.3033 & 9.045 \\
2302     & 1234pent & F       & B        & C        & $-$10.3181 & $-$4.3767 & 5.941 \\
2926     & C$_{35}$H$_{36}$ & F & C      & C        & $-$10.2891 & $-$2.1723 & 8.117 \\
797      & tri      & F       & B        & C        & $-$10.2849 & $-$2.8059 & 7.479 \\
1115     & 1-2-3tet & F       & B        & Si       & $-$10.2796 & $-$2.9255 & 7.354 
\end{tabular}
\end{table}

\begin{table}[h]
 \makeatletter 
 \renewcommand{\thetable}{S\@arabic\c@table}
 \makeatother
\label{tab:HighHOMO}
\caption{Configurations of the ten ND5k structures with the highest HOMO energies: ND5k index, base structure (ND), surface species, dopants (D1, D2), HOMO, LUMO and gap energies (all in eV).}
\medskip
\begin{tabular}{llllllll}
Index & ND & Surface & D1 & D2 & E(HOMO) & E(LUMO) & E(gap)   \\
\hline
3921     & C$_{74}$H$_{64}$  & H       & P        & P        & 0.0327  & 0.5572 & 0.525 \\
2709     & C$_{35}$H$_{36}$  & H       & N        & P        & $-$0.0974 & 0.522  & 0.619 \\
4599     & C$_{104}$H$_{78}$ & NH$_2$     & P        & P        & $-$0.2768 & 1.1598 & 1.437 \\
2711     & C$_{35}$H$_{36}$  & H       & N        & P        & $-$0.3804 & 0.4121 & 0.792 \\
4823     & C$_{109}$H$_{80}$ & H       & P        & P        & $-$0.3808 & 0.7989 & 1.18  \\
4521     & C$_{104}$H$_{78}$ & H       & P        & P        & $-$0.383  & 0.8382 & 1.221 \\
3016     & C$_{35}$H$_{36}$  & H       & P        & P        & $-$0.4998 & 0.6747 & 1.174 \\
3397     & C$_{53}$H$_{48}$  & NH$_2$/H   & P        & P        & $-$0.5141 & 0.9086 & 1.423 \\
4809     & C$_{109}$H$_{80}$ & H       & C        & P        & $-$0.54   & 1.1041 & 1.644 \\
4602     & C$_{104}$H$_{78}$ & NH$_2$/H   & P        & P        & $-$0.5402 & 1.01   & 1.55 
\end{tabular}
\end{table}

\begin{table}[h]
 \makeatletter 
 \renewcommand{\thetable}{S\@arabic\c@table}
 \makeatother
\label{tab:LowLUMO}
\caption{Configurations of the ten ND5k structures with the lowest LUMO energies: ND5k index, base structure (ND), surface species, dopants (D1, D2), HOMO, LUMO and gap energies (all in eV).}
\medskip
\begin{tabular}{llllllll}
Index & ND & Surface & D1 & D2 & E(HOMO) & E(LUMO) & E(gap)   \\
\hline
5004     & C$_{109}$H$_{80}$ & F       & B        & B        & $-$9.4822 & $-$8.6369 & 0.845 \\
4402     & C$_{88}$H$_{80}$  & F       & B        & B        & $-$9.0967 & $-$8.5691 & 0.528 \\
4703     & C$_{104}$H$_{78}$ & F       & B        & B        & $-$9.6228 & $-$8.2213 & 1.402 \\
4102     & C$_{74}$H$_{64}$  & F       & B        & B        & $-$9.8487 & $-$7.8715 & 1.977 \\
5007     & C$_{109}$H$_{80}$ & F/H     & B        & B        & $-$7.7438 & $-$6.3962 & 1.348 \\
4405     & C$_{88}$H$_{80}$  & F/H     & B        & B        & $-$7.9938 & $-$6.3828 & 1.611 \\
195      & ad      & F       & B        & B        & $-$9.447  & $-$5.5824 & 3.865 \\
4706     & C$_{104}$H$_{78}$ & F/H     & B        & B        & $-$7.8016 & $-$5.5813 & 2.22  \\
3855     & C$_{68}$H$_{64}$  & F       & N        & P        & $-$9.0876 & $-$5.4142 & 3.673 \\
4618     & C$_{104}$H$_{78}$ & OH      & B        & B        & $-$5.9422 & $-$5.3277 & 0.614
\end{tabular}
\end{table}

\begin{table}[h]
 \makeatletter 
 \renewcommand{\thetable}{S\@arabic\c@table}
 \makeatother
\label{tab:HighLUMO}
\caption{Configurations of the ten ND5k structures with the highest LUMO energies: ND5k index, base structure (ND), surface species, dopants (D1, D2), HOMO, LUMO and gap energies (all in eV).}
\medskip
\begin{tabular}{llllllll}
Index & ND & Surface & D1 & D2 & E(HOMO) & E(LUMO) & E(gap)   \\
\hline
674      & tri      & NH$_2$     & N        & P        & $-$5.0044 & 1.5662 & 6.571 \\
368      & di       & NH$_2$     & N        & N        & $-$5.6194 & 1.5118 & 7.131 \\
2152     & 1234pent & NH$_2$     & C        & C        & $-$4.9126 & 1.5059 & 6.418 \\
2160     & 1234pent & NH$_2$     & C        & P        & $-$5.0175 & 1.4982 & 6.516 \\
660      & tri      & NH$_2$/H   & C        & P        & $-$5.5096 & 1.4875 & 6.997 \\
68       & ad       & NH$_2$/H   & C        & C        & $-$5.7443 & 1.4779 & 7.222 \\
3057     & C$_{35}$H$_{36}$ & NH2 & C       & N        & $-$4.1135 & 1.4734 & 5.587 \\
22       & ad       & H       & N        & N        & $-$5.5275 & 1.4721 & 7.0   \\
1573     & 123tet   & NH$_2$/H   & N        & N        & $-$5.3013 & 1.4698 & 6.771 \\
86       & ad       & NH$_2$/H   & N        & N        & $-$5.5512 & 1.4659 & 7.017
\end{tabular}
\end{table}

\begin{table}[h]
 \makeatletter 
 \renewcommand{\thetable}{S\@arabic\c@table}
 \makeatother
\label{tab:Lowgap}
\caption{Configurations of the ten ND5k structures with the lowest gap energies: ND5k index, base structure (ND), surface species, dopants (D1, D2), HOMO, LUMO and gap energies (all in eV).}
\medskip
\begin{tabular}{llllllll}
Index & ND & Surface & D1 & D2 & E(HOMO) & E(LUMO) & E(gap)   \\
\hline
3929      & C$_{74}$H$_{64}$  & NH$_2$  & B       & B       & $-$3.0854 & $-$2.6493 & 0.436 \\
4230      & C$_{88}$H$_{80}$  & NH$_2$  & B       & B       & $-$3.1821 & $-$2.6925 & 0.49  \\
3628      & C$_{68}$H$_{64}$  & NH$_2$  & B       & B       & $-$3.04   & $-$2.5285 & 0.511 \\
3921      & C$_{74}$H$_{64}$  & H       & P       & P       & 0.0327  & 0.5572  & 0.525 \\
4402      & C$_{88}$H$_{80}$  & F       & B       & B       & $-$9.0967 & $-$8.5691 & 0.528 \\
4618      & C$_{104}$H$_{78}$ & OH      & B       & B       & $-$5.9422 & $-$5.3277 & 0.614 \\
2709      & C$_{35}$H$_{36}$  & H       & N       & P       & $-$0.0974 & 0.522   & 0.619 \\
2711      & C$_{35}$H$_{36}$  & H       & N       & P       & $-$0.3804 & 0.4121  & 0.792 \\
4616      & C$_{104}$H$_{78}$ & OH      & B       & B       & $-$5.8127 & $-$4.9751 & 0.838 \\
5004      & C$_{109}$H$_{80}$ & F       & B       & B       & $-$9.4822 & $-$8.6369 & 0.845
\end{tabular}
\end{table}

\begin{table}[h]
 \makeatletter 
 \renewcommand{\thetable}{S\@arabic\c@table}
 \makeatother
\label{tab:Highgap}
\caption{Configurations of the ten ND5k structures with the highest gap energies: ND5k index, base structure (ND), surface species, dopants (D1, D2), HOMO, LUMO and gap energies (all in eV).}
\medskip
\begin{tabular}{llllllll}
Index & ND & Surface & D1 & D2 & E(HOMO) & E(LUMO) & E(gap)   \\
\hline
224  & ad       & F & C & C & $-$9.3729  & 0.6944  & 10.067 \\
519  & di       & F & C & C & $-$9.1334  & 0.7428  & 9.876  \\
820  & tri      & F & C & C & $-$8.9487  & 0.6994  & 9.648  \\
2325 & 1234pent & F & C & C & $-$9.2981  & 0.2987  & 9.597  \\
1100 & 1-2-3tet & F & B & C & $-$8.6555  & 0.8666  & 9.522  \\
1422 & 121tet   & F & C & C & $-$8.6758  & 0.7715  & 9.447  \\
1121 & 1-2-3tet & F & C & C & $-$8.3746  & 1.0361  & 9.411  \\
2324 & 1234pent & F & C & C & $-$10.3749 & $-$0.9683 & 9.407  \\
2024 & 1212pent & F & C & C & $-$9.1963  & 0.2032  & 9.4    \\
228  & ad       & F & C & N & $-$8.5554  & 0.8367  & 9.392
\end{tabular}
\end{table}

\clearpage

From tables S1 to S6 one can deduce trends of structural patterns and their effect on the nanodiamonds' energetic properties. Low HOMO energies are obtained from small diamondoids with full F coverage and no or little B/Si doping (table S1). In contrast, high HOMO energys are obtained from larger NDs with H- or NH$_2$ doping and heavy n-type doping, especially with P (table S2). The NDs with the lowest LUMO energies are large structures with F-termination and heavy B-doping, however, there are two outliers present (table S3): In the small B-B doped fluorinated adamantane, two CF groups are replaced by B, resulting in a structure with many CF$_2$ groups, some CF groups and double B doping. The relatively large number of CF$_2$ groups is likely the reason for the unusually low LUMO energy of this adamantane. For the F-terminated, N-P doped C$_{68}$H$_{64}$, the special LUMO may originate from the structure's nitrogen vacancy (NV) center. The highest LUMO energies are obtained mainly from small, NH$_2$-terminated and doubly n-type doped diamondoids (table S4). The structures with the smallest gaps are rather diverse, ranging from medium to large sizes, being either doubly p-type doped or doubly n-type doped and having all sorts of surface terminations in various combintaions with their doping patterns (table S5). Most likely, the introduction of either low lying unoccupied or high lying occupied states by heavy doping is the main ingredient towards a small gap. Finally, undoped F-terminated diamondoids make up most of the structures with the largest gaps, as a result of concurrent F-termination and the quantum confinement effect (table S6). All of these findings are generally in agreement with previous studies\cite{willey_molecular_2005, teunissen_tuning_2017} and can be regarded as design suggestions for any suitable application.

\clearpage

\section{Details of the Machine Learning Setups}\label{SecML}

We obtained the following hyperparameters for the SOAP KRR: 6 \r{A} interaction cutoff, $n_{max}$ (radial) $=$ 6, $l_{max}$ (angular) $=$ 9, gaussian sigma const. $=$ 0.4, kernel zeta $=$ 3.\cite{musil2018librascal}

For ML with the enn-s2s, molecular graphs were created as radius graphs from the DFTB-optimized structures, i.e., each atom node was connected to all of its neighbors within a given cutoff radius. We selected the following inputs for the node feature vectors: one-hot element type, atomic number, valency, atomic radius, Pauling electronegativity, electron affinity and 1st ionization energy, all normalized to values $\leq$ 1. The edge features were chosen to be bond length and (cutoff radius $-$ bond length). Leaky ReLU was used as the nonlinear function in the dense NNs and the ADAM optimizer was employed for minimizing the mean squared error of the network predictions. 

In the enn-s2s architecture, first, the node features are transformed via a dense NN. Subsequently, the NNConv graph convolutional layer\cite{Fey/Lenssen/2019} is applied $n$ times, including two dense NN layers and batch normalization. A set2set operation is performed on the transformed graphs to obtain fixed-size vectors, and a dropout layer is applied for regularization. Finally, these vectors are passed to a 3-layer dense neural network with decreasing layer size which finally outputs the target values. A similar architecture was recently used for predicting frontier orbital energies of diverse medium-sized organic molecules (OE62 dataset).\cite{stuke_atomic_2020, rahaman_deep_2020}

We obtained the following enn-s2s hyperparameters (identical for HOMO and LUMO energy predictions): 3.5 \r{A} cutoff radius, 0.001 learning rate, n$_{\mathrm{pre}}$ $=$ 3, p1 $=$ 64, p2 $=$ 64, p$_{\mathrm{dropout}}$ $=$ 0.3, n$_{\mathrm{layers}}$ $=$ 4. The parameter n$_{\mathrm{pre}}$ defnies the number of layers in the dense NN that acts on the initial node embeddings, the parameters p1 and p2 control the size of the network layers, p$_{\mathrm{dropout}}$ defines the dropout probability, and n$_{\mathrm{layers}}$ defines the number of NNConv operations before the set2set evaluation. The batch size was fixed to 8 for all training runs due to memory restrictions. We found that small changes of the hyperparameters did not have large effects on the enn-s2s performance. The enn-s2s and its variants used a batchsize of 8 for training.

For the SOAP-enn-s2s and SOAP-PCA-enn-s2s approach, we used the following SOAP hyperparameters: 3.5 \r{A} interaction cutoff, $n_{max}$ (radial) $=$ 3, $l_{max}$ (angular) $=$ 3, gaussian sigma const. $=$ 0.5 and a soft (polynomial) cutoff of the form proposed by Caro.\cite{caro_optimizing_2019} After another short hyperparameter optimization run of the SOAP-(PCA)-enn-s2s, only the number of layers in the initial dense NN was increased to n$_{\mathrm{pre}}$ $=$ 5.

The SchNet hyperparameters that we found to be optimal for training with ND5k (both HOMO and LUMO energies) are close to its default hyperparameters, with only the cutoff radius and num\_gaussians being modified: 3.5 \r{A} cutoff, hidden channels $=$ 64, num\_filters $=$ 128, num\_interactions $=$ 6, num\_gaussians $=$ 70, batchsize $=$ 10.\cite{Fey/Lenssen/2019} 

The PaiNN GNN was trained using the following hyperparameters: 7.0 \r{A} cutoff, n\_atom\_basis $=$ 150, n\_interactions $=$ 3, n\_rbf $=$ 25, max\_z $=$ 100, batchsize $=$ 5.\cite{schutt_equivariant_2021}

The enn-s2s and SOAP-PCA-enn-s2s architectures were further fine-tuned for learning frontier orbital energies on QM9 and OE62 using a small grid search. Final hyperparameters are obtained as follows. QM9: 5.0 \r{A} cutoff radius, 0.001 learning rate, n$_{\mathrm{pre}}$ $=$ 5, p1 $=$ 96, p2 $=$ 96, p$_{\mathrm{dropout}}$ $=$ 0.3, n$_{\mathrm{layers}}$ $=$ 4. SOAP parameters for the SOAP-PCA variant were 3.5 \r{A} cutoff radius, $n_{max}$ (radial) $=$ 2, $l_{max}$ (angular) $=$ 2 and a soft (polynomial) cutoff of the form proposed by Caro.\cite{caro_optimizing_2019}
OE62: 3.0 \r{A} cutoff radius, 0.001 learning rate, n$_{\mathrm{pre}}$ $=$ 3, p1 $=$ 96, p2 $=$ 96, p$_{\mathrm{dropout}}$ $=$ 0.3, n$_{\mathrm{layers}}$ $=$ 4. SOAP parameters for the SOAP-PCA variant were 3.0 \r{A} cutoff radius, $n_{max}$ (radial) $=$ 3, $l_{max}$ (angular) $=$ 3 and a hard cutoff.
For training, both datasets were randomly split into train, validation and test subsets containing 70 \%, 15 \% and 15 \% of the original samples, respectively. Training was performed as described before, however, due to the larger sizes of the datasets, training was limited to 150 epochs and the learning rate scheduler was set to reduce the learning rate by a factor of 0.7 every 15 epochs if learning stagnates. Additionally, the plain SOAP method was not employed as the larger memory requirements of this method were computationally unfeasible, and the SOAP PCA method has already been shown to yield better results for the ND5k dataset. Furthermore, as a full PCA was not feasible due to memory constraints, so the memory-efficient incremental PCA algorithm proposed by Ross et al. was used instead.\cite{ross_incremental_2008} In contrast to ND5k, where the retained PCA dimensionality could be chosen to retain a certain fraction of variance, this method demands to fix the retained PCA dimension up front. To account for the fact that the OE62 dataset contains more elements than ND5k, the retained dimensionality was fixed to 200 for OE62, and a dimensionality of 100 was chosen for QM9, which contains only five atom types. Training results for these datasets were averaged over three independent runs.

\clearpage

\section{Correlations between nanodiamond size and HOMO/LUMO energy}\label{SecML}

Figures \ref{fig:correlationHOMO} and \ref{fig:correlationLUMO} show scatter plots of the HOMO and LUMO energies vs. the number of atoms for all structures in the ND5k dataset (blue) and in the ND5k-lt set (orange) that contains 24 larger nanodiamonds. The red lines indicate the linear fits within the ND5k dataset. As can be seen, especially for the HOMO energies, the correlation that is indicated by the linear fit breaks down for the ND5k-lt structures. These structures mostly have lower frontier orbital energies than anticipated by the fit.

\begin{figure}[H]
 \makeatletter 
 \renewcommand{\thefigure}{S\@arabic\c@figure}
 \makeatother
 \raggedright
 \includegraphics[width = 0.6\textwidth]{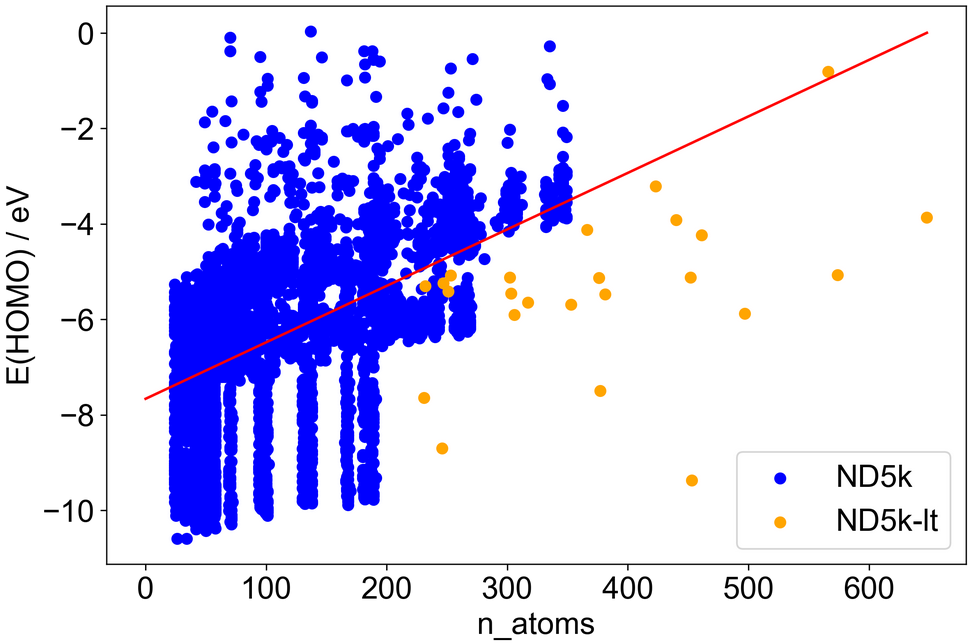}
    \caption{Scatter plot of HOMO energy and number of atoms of the ND5k structures (blue) and the large nanodiamonds (orange). The red line indicates the linear fit on the ND5k structures.}
    \label{fig:correlationHOMO}
\end{figure} 

\begin{figure}[H]
 \makeatletter 
 \renewcommand{\thefigure}{S\@arabic\c@figure}
 \makeatother
 \raggedright
 \includegraphics[width = 0.6\textwidth]{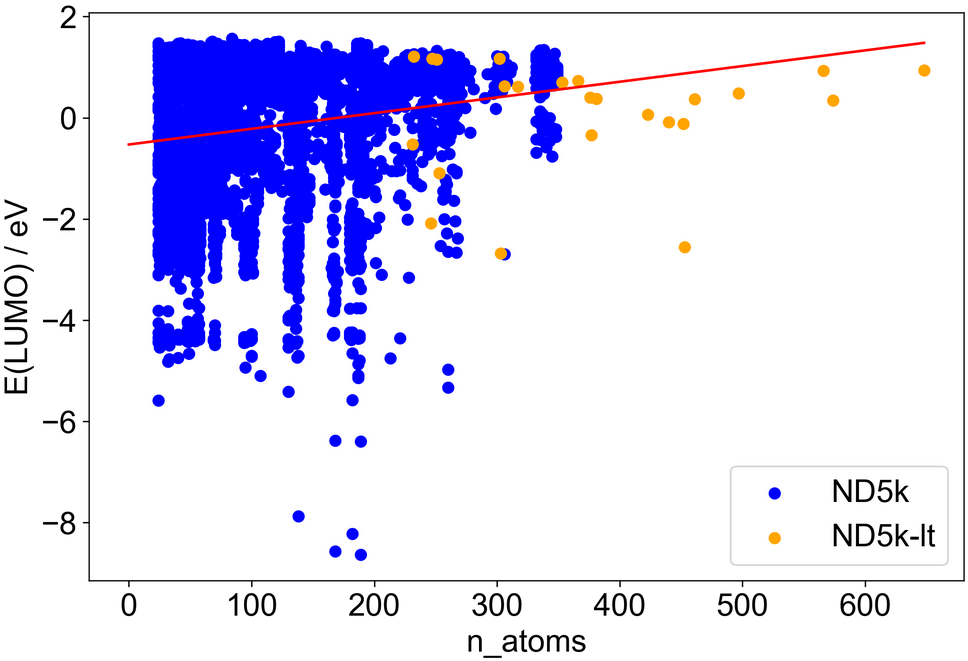}
    \caption{Scatter plot of LUMO energy and number of atoms of the ND5k structures (blue) and the large nanodiamonds (orange). The red line indicates the linear fit on the ND5k structures.}
    \label{fig:correlationLUMO}
\end{figure}

\end{justify}

\bibliography{references}
\bibliographystyle{unsrt}